\documentclass[aps,pre,showpacs,preprint,superscriptaddress]{revtex4}
\usepackage{graphicx}
\usepackage{subfigure}
\usepackage{amsmath}
\usepackage{ulem}
\usepackage{bm}
\usepackage{amssymb}
\usepackage{color}
\usepackage{epstopdf}
\usepackage{epsfig,dsfont,multirow}

\begin{document}
\title{Angular distributions of nonlinear Thomson scattering in combining field with a general elliptically polarized laser and a background magnetic field}
\author{Hong Xiao}
\affiliation{Key Laboratory of Beam Technology of the Ministry of Education, and College of Nuclear Science and Technology, Beijing Normal University, Beijing 100875, China}
\author{Cui-Wen Zhang}
\affiliation{Key Laboratory of Beam Technology of the Ministry of Education, and College of Nuclear Science and Technology, Beijing Normal University, Beijing 100875, China}
\author{Hai-Bo Sang}
\affiliation{Department of Physics, Beijing Normal University, Beijing 100875, China}
\author{Bai-Song Xie}
\email[Author to whom correspondence should be addressed. Electronic mail: ]{bsxie@bnu.edu.cn}
\affiliation{Key Laboratory of Beam Technology of the Ministry of Education, and College of Nuclear Science and Technology, Beijing Normal University, Beijing 100875, China}
\affiliation{Institute of Radiation Technology, Beijing Academy of Science and Technology, Beijing 100875, China}
\date{\today}
\begin{abstract}
Nonlinear Thomson scattering of an electron motion in a combining field constituted by an elliptically polarized laser and a background magnetic field is investigated. The dependence of the electron trajectories, the fundamental frequency, the maximum radiation power in spatial distribution and corresponding spatial angle on ellipticity are obtained. In addition, we find that the angular distributions of scattering spectra with respect to the azimuthal angle exhibits the symmetry no matter what the order of harmonics, the laser intensity, the magnetic resonance parameter and the initial axial momentum are. Meanwhile, the polar angle distribution of the spectra approaches more and more the laser propagation direction with the laser intensity, the magnetic resonance parameter and the initial axial momentum. The maximum radiated power increases and the corresponding polar angle decreases. The optimal angle for the maximum radiated power per unit of solid, the corresponding photon number and the photons brightness can be obtained, which implies that the high quality XUV or/and x-ray can be generated by the studied scheme when the suitable parameters are chosen.
\end{abstract}
\maketitle

\section{Introduction}

A great increase in the laser intensity by the chirped pulse amplification (CPA) \cite{Maine-24-398} technique in the past decades intrigues more and more research on the ultrashort ultraintense lasers interacting with matter. Among them the nonlinear Thomson scattering (NTS) has become an important research area due to its great value and application prospect to generate bright and ultrashort x-ray source \cite{Sprangle-72-5032,PRE-48-3003,Kim-341-351,Kotaki-455-166,Brown-7-060703,Anderson-78-891,Tomassini-80-419,PRL-96-014802} or/and $\gamma$-ray pulse \cite{Taira-7-5018,PRA-98-052130,Chen-4-024401}. Furthermore, the attosecond pulse \cite{POP-13-013106,PRE-72-066501,Wang-31-015301}, higher order of harmonics \cite{Venkat-22-084401} and the optical vortices \cite{Taira-860-45} can also be generated by NTS.

It is well known that obvious difference of spatial radiation features when parameters of the incident laser pulse are changing, such as polarization state, initial phase, beam waist and pulse width.

For linearly polarized laser field, Chen \textit{et al.} \cite{Nat-396-653} experimentally measured the angular radiation profiles of the second and third harmonics. In 2009, the spatial characteristics of the Thomson scattering were theoretically studied in a linearly polarized laser field by Lan \textit{et al.}, which shows that the spatial distributions depends sensitively on the initial phase \cite{Lan-9-143}. Vais and Yu studied the influence of the beam waist on spatial characteristics \cite{PPR-42-818,Yu-30-045301}. Zhang found the four-fold and two-fold rotational symmetry pattern for the space radiation characteristics under different laser intensities \cite{Zhang-137-262} and Li discovered the bifoliate radiation pattern (BRP) \cite{Li-183-813}. Ruijter theoretically and numerically showed the dependency of the spectrum on the intensity of the laser and the carrier envelope phase \cite{Ruijter-11-528}.

For circularly polarized laser field \cite{PRE-68-056501,Boca-83-055404,Tian-45-1125,Zhao-36-074101,Wu-30-115301,Wang-18-015303,Wang-31-015301,Liu-74-7,Chen-31-075401,Wang-53-229,Shi-32-015401}, the researches indicated that the initial phase has a great influence on the spatial distribution \cite{PRE-68-056501,Tian-45-1125,Shi-32-015401}. Wu \textit{et al.} found that the radiation was significantly affected by beam waist radius \cite{Wu-30-115301}. With the increasing of laser intensity, the peak radiation power increases, polar angle decreases and the direction of radiation approaches the direction of laser propagation \cite{Zhao-36-074101,Wang-18-015303,Chen-31-075401,Wang-53-229}.

Moreover, He \textit{et al.} indicated the different effects by varying the initial phase for the circularly and linearly polarized laser fields in the combination of an intense laser field and a strong uniform magnetic field \cite{PRE-68-056501}. For elliptically polarized laser field, Shi \textit{et al.} researched the spatial radiation distribution with varied incident pulse widths \cite{Shi-31-105401}. Fruhling \textit{et al.} reported experiment results of the radiated polarization states by elliptically polarized laser \cite{PRA-104-053519}.

In recent years, the properties of Thomson back-scattering spectra have been explored concretely in combining magnetic and laser field \cite{Yu-30-045301,Ruijter-11-528} and the influences of the relevant parameters such as the laser intensity, the initial electron's axial momentum are found \cite{PRA-94-052102,EPL-117-44002,EPL-126-34001,EPL-125-64002}. On the other hand, many studies on spatial radiation characteristics of NTS are limited only for the single linearly or/and circularly polarized laser field. To our knowledge, the investigation of complete spatial radiation characteristics of single electron interacting with the combination of an intense elliptically polarized laser field and a strong magnetic field are missed, therefore, it is worth to study it in detail since
some new features should be useful to get the high quality x-ray source by the NTS in this situation.

In this paper, we study the NTS in the combining field by an analytical treatment and numerical calculation. The relation of the maximum radiated power and corresponding spatial angle to ellipticity $\alpha$ with different harmonic order number $m$, laser intensity $I_{0}$, magnetic resonance parameter $n$ and initial axial momentum $p_{z0}$ are obtained. In $zx$ plane, when azimuthal angle $\varphi=0$, the elliptically polarized laser is better than linearly and circularly polarized ones under some parameters. In addition, we get the angular distribution with respect to azimuthal angle and polar angle. And the various relationship between radiation characteristics and parameters are revealed. The results show that high quality XUV or/and x-ray can be produced via NTS in the combining field. In the optimal emission  angles, the number of the emitted photons is approaching $10^{12}$-$10^{15}$ and the photons brightness can reach $10^{20}\mathrm{photons}/\mathrm{s}~\mathrm{mm}^{2}\mathrm{mrad}^{2} 0.1 \%\mathrm{BW}$.

The paper is organized as follows. In Sec.II, the momentum and trajectory equations of the electron in the combining field are obtained. And the radiation power per unit of solid angle in the emission direction is given. In Sec.III, the dependence of various physical quantities involved in NTS on ellipticity are examined such as the electron trajectories, the fundamental frequency, the maximum radiation power in spatial distribution and the corresponding spatial angle and so on. In Sec.IV, the radiation distribution of the NTS with respect to the azimuthal angle and polar angle is obtained. Meanwhile the angular distributions of NTS are presented with different parameters. In particular, the largest radiated power, the number of photons and the photons brightness are given in an optimum emission  direction. Finally the main conclusions and some discussions are given briefly in Sec. V.

\section{basic equations}

\begin{figure}[htbp]\suppressfloats
\includegraphics[width=15cm]{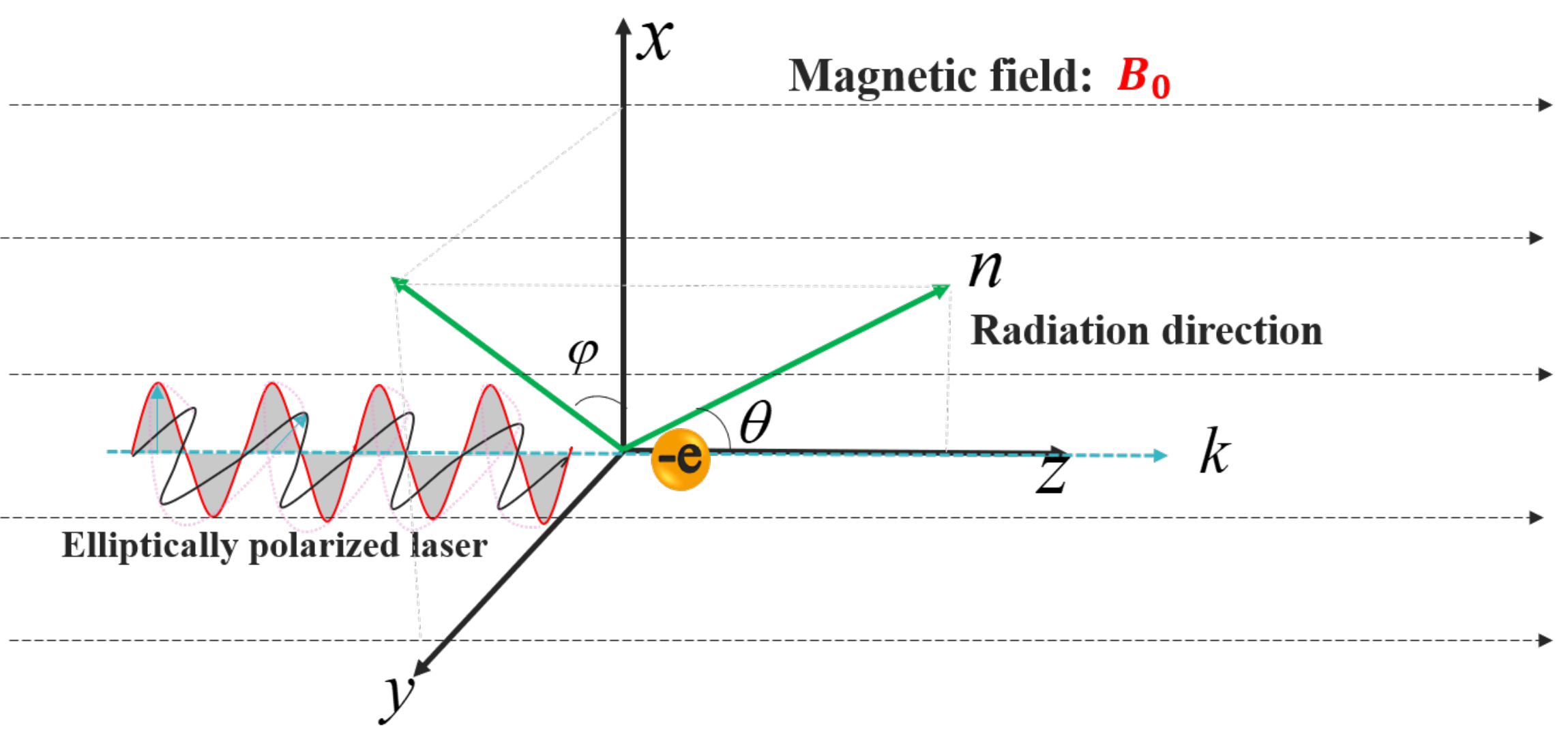}
\caption{\label{Fig1}(colour online) A schematic illustration of physical model.}
\end{figure}
Let us consider the spatial angular distributions of NTS that the electron (with mass $m$ and charge $-e$) moves in the combining field. As shown in Fig.\ref{Fig1}, we assume that the amplitude of external magnetic field is $B_{0}$, and the laser field is left-hand elliptically polarized with vector potential amplitude $A_{0}$, frequency $\omega_{0}$ and ellipticity $\alpha$. The laser propagation direction and the magnetic direction are both along the $+z$ direction. By denoting the phase of the laser field as $\eta={\omega}_{0}t-\textit{\textbf{k}}\cdot\textit{\textbf{r}}$, where $\textit{\textbf{k}}$ and $\textit{\textbf{r}}$ are the laser wave vector and electron displacement vector respectively. The combining field can be expressed by the total vector potential:

\begin{equation}
\label{eq1}\textit{\textbf{A}}=
\frac{{A}_{0}}{\sqrt{1+\alpha^2}}\left [-\sin {\eta}\textit{\textbf{i}}+\alpha\cos{\eta}\textit{\textbf{j}}\right ]+{B}_{0}x\textit{\textbf{j}}.
\end{equation}
From Eq.(\ref{eq1}), the corresponding electric field $\textit{\textbf{E}}$ and magnetic field $\textit{\textbf{B}}$ can be obtained
\begin{equation}
\label{eq2}\textit{\textbf{E}}= -\frac{1}{c} \frac{\partial \textit{\textbf{A}}}{\partial t},
\end{equation}
\begin{equation}
\label{eq3}\textit{\textbf{B}}=\nabla \times \textit{\textbf{A}}.
\end{equation}
Then, the dynamics of electron will be studied according to the momentum-energy evolving equations:
\begin{equation}
\label{eq4}d\textit{\textbf{p}}/dt=-e(\textit{\textbf{E}}+\boldsymbol{\beta}\times \textit{\textbf{B}}),
\end{equation}
and
\begin{equation}
\label{eq5}d\left(\gamma mc^2\right)/dt=-ec\boldsymbol{\beta}\cdot\textit{\textbf{E}},
\end{equation}
where $\textit{\textbf{p}}$ is the electron's momentum, $\boldsymbol{\beta}$ is the electron's velocity, $\gamma=\left(1-\beta^2\right)^{-1/2}$ is the electron relativistic factor. Note that in order to simplify the calculations, we have normalized time by $1/\omega_{0}$, distance by $1/k_{0}$, velocity by $c$, momentum by $mc$, the electric field by $e{E}_{0}/m\omega_{0}c$, and the magnetic field by $e{B}_{0}/m\omega_{0}c$.

By substituting Eqs.(\ref{eq2}) and (\ref{eq3}) into the Eq.(\ref{eq4}), we get
\begin{equation}
\label{eq6}d^2 p_{x} /d \eta ^2+{\omega _{b}}^2 p_{x}= (\omega_{b}\alpha +1)a\sin{\eta},
\end{equation}
\begin{equation}
\label{eq7}d^2 p_{y} /d \eta ^2+{\omega _{b}}^2 p_{y}=-(\alpha+\omega_{b})a\cos{\eta},
\end{equation}
where $a=eA_{0}/(\sqrt{1+\alpha^2}mc^{2})$ is the normalized vector potential amplitude, $\omega_{b}= b/ \varsigma$ is the cyclotron frequency of the electron motion in the combining field and $b$ is the magnetic field strength. We assume that the initial value of the phase $\eta$ is $\eta_{in}=-z_{in}$. Since an electron moves in a constant amplitude elliptically polarized laser field, the constant of the motion can be obtained as $\varsigma =\gamma-p_{z}=\gamma_{0}-p_{z0}=\sqrt{1+p_{z0}^{2}}-p_{z0}$ according to Eqs.(\ref{eq4}) and (\ref{eq5}), where $p_{z0}=\frac{1}{2\varsigma}-\frac{\varsigma}{2}$ is the initial momentum in the $z$ direction.

By solving Eqs.(\ref{eq6}) and (\ref{eq7}), the electron momentum and the trajectories via $d\textit{\textbf{r}}/d\eta=\textit{\textbf{p}}/\varsigma$ can be obtained as
\begin{equation}
\label{eq8}p_{x}=na\left \{\varepsilon_{1}\sin{\eta}-\varepsilon_{1}\cos{\left[\omega_{b}\left( \eta-\eta_{in}\right)\right]}\sin{\eta_{in}}-\varepsilon_{2}\sin{\left[\omega_{b}\left( \eta-\eta_{in}\right)\right]}\cos{\eta_{in}} \right \},
\end{equation}
\begin{equation}
\label{eq9}p_{y}=na\left \{-\varepsilon_{2}\cos{\eta}-\varepsilon_{2}\cos{\left[\omega_{b}\left( \eta-\eta_{in}\right)\right]}\cos{\eta_{in}}-\varepsilon_{1}\sin{\left[\omega_{b}\left( \eta-\eta_{in}\right)\right]}\sin{\eta_{in}} \right \},
\end{equation}
\begin{equation}
\begin{aligned}
\label{eq10}p_{z}=&\frac{n^2a^2}{2\varsigma}
\left \{{\varepsilon_{1}^2\left(\sin^2{\eta}+\sin^2{\eta_{in}}\right)
+\varepsilon_{2}^2\left(\cos^2{\eta}+\cos^2{\eta_{in}}\right)}\right. \\& \left.{-2\cos{\left[\omega_{b}\left( \eta-\eta_{in}\right)\right]}\left(\varepsilon_{1}^2\sin{\eta}\sin{\eta_{in}}
 +\varepsilon_{2}^2\cos{\eta}\cos{\eta_{in}}\right)}\right. \\& \left.{-2\varepsilon_{1}\varepsilon_{2}\sin{\left[\omega_{b}\left(\eta-\eta_{in}\right)\right]}\sin{\left( \eta-\eta_{in}\right)}}\right \}+\frac{1}{2\varsigma}-\frac{\varsigma}{2},
\end{aligned}
\end{equation}
and
\begin{equation}
\begin{aligned}
\label{eq11}x(\eta)=&\frac{na}{\varsigma} \left \{{-\varepsilon_{1}\cos{\eta}-\frac{\varepsilon_{1}}{\omega_{b}}\sin{\left[\omega_{b}\left( \eta-\eta_{in}\right)\right]}\sin{\eta_{in}} }\right. \\& \left.{
+\frac{\varepsilon_{2}}{\omega_{b}}\cos{\left[\omega_{b}\left(\eta-\eta_{in}\right)\right]}\cos{\eta_{in}}
+\left(\varepsilon_{1}-\frac{\varepsilon_{2}}{\omega_{b}}\right)\cos{\eta_{in}}}\right \},
\end{aligned}
\end{equation}
\begin{equation}
\begin{aligned}
\label{eq12}y(\eta)=&\frac{na}{\varsigma} \left \{{-\varepsilon_{2}\sin{\eta}+\frac{\varepsilon_{1}}{\omega_{b}}\cos{\left[\omega_{b}\left( \eta-\eta_{in}\right)\right]}\sin{\eta_{in}} }\right. \\& \left.{
+\frac{\varepsilon_{2}}{\omega_{b}}\sin{\left[\omega_{b}\left(\eta-\eta_{in}\right)\right]}\cos{\eta_{in}}
+\left(\varepsilon_{2}-\frac{\varepsilon_{1}}{\omega_{b}}\right)\sin{\eta_{in}}}\right \},
\end{aligned}
\end{equation}
\begin{equation}
\begin{aligned}
\label{eq13}z(\eta)=& \frac{(n a)^{2}}{2  \varsigma^{2}}\left\{-\varepsilon_{1}^{2}\left[\frac{\sin 2 \eta}{4}-\left(\sin ^{2} \eta_{i n}+\frac{1}{2}\right)\left(\eta-\eta_{i n}\right)\right]\right.\\
&+\varepsilon_{2}^{2}\left[\frac{\sin 2 \eta}{4}+\left(\cos ^{2} \eta_{i n}+\frac{1}{2}\right)\left(\eta-\eta_{i n}\right)\right] \\
&+\frac{1}{2}\left(\varepsilon_{1}^{2}-\varepsilon_{2}^{2}\right)\left\{\frac{\sin \left[\left(\omega_{b}-1\right) \eta-\left(\omega_{b}+1\right) \eta_{i n}\right]}{\omega_{b}-1}\right.\\
&\left.+\frac{\sin \left[\left(\omega_{b}+1\right) \eta-\left(\omega_{b}-1\right) \eta_{i n}\right]}{\omega_{b}+1}\right\} \\
&-\left[\frac{1}{2}\left(\varepsilon_{1}^{2}+\varepsilon_{2}^{2}\right)+\varepsilon_{1} \varepsilon_{2}\right] \frac{\sin \left[\left(\omega_{b}-1\right)\left(\eta-\eta_{i n}\right)\right]}{\omega_{b}-1} \\
&\left.-\left[\frac{1}{2}\left(\varepsilon_{1}^{2}+\varepsilon_{2}^{2}\right)-\varepsilon_{1} \varepsilon_{2}\right] \frac{\sin \left[\left(\omega_{b}+1\right)\left(\eta-\eta_{i n}\right)\right]}{\omega_{b}+1}\right\}\\
&+\left(\frac{1}{2 \varsigma^{2}}-\frac{1}{2}\right)\left(\eta-\eta_{i n}\right),
\end{aligned}
\end{equation}
respectively, where $\varepsilon_{1}=\left(\omega_{b} \alpha+1\right) /\left(\omega_{b}+1\right)$ and $\varepsilon_{2}=$ $\left(\alpha+\omega_{b}\right) /\left(\omega_{b}+1\right)$. We just consider the high frequencies, and $n$ is the magnetic resonance parameter as $n=1/\left (\omega_{b}-1\right)$, refer to Ref.\cite{Li-95-161105}. So, the strength of external magnetic field  $B_{0}=(1+1/n)(\sqrt{1+p_{z0}^{2}}-p_{z0})m\omega_{0}c/e\approx[(1+1/n)(\sqrt{1+p_{z0}^{2}}-p_{z0})/\lambda[\mathrm{\mu m}]]\times100\mathrm{MG}$, which is obviously related to $n$ and $p_{z0}$. We have $B_{0}\approx(1+1/n)\times\mathrm{100MG}$ when $p_{z0}\ll1$ and $B_{0}\approx[(1+1/n)/2p_{z0}]\times\mathrm{100MG}$ when $p_{z0}\gg1$.

The angular distributions of the emission power detected far away from the electron in the direction $\textit{\textbf{n}}$ (not to be confused with the magnetic resonance parameter $n$) can be calculated as \cite{Jakson}:
\begin{equation}
\label{eq14}\frac{d^2 I }{d\Omega d\omega }=\frac{e^2\omega^2}{4\pi ^2c}\left|\textit{\textbf{n}}\times \left[\textit{\textbf{n}}\times\textit{\textbf{F}}(\omega) \right ]\right|^2,
\end{equation}
with a function of
\begin{equation}
\label{eq15}\textit{\textbf{F}}(\omega)=\frac{1}{\varsigma}\int_{-\infty }^{+\infty}d\eta\textit{\textbf{p}}(\eta)\exp{\left \{ i\omega\left [ \eta-\textit{\textbf{n}}\cdot\textit{\textbf{r}}(\eta)+z(\eta)\right ] \right \}},
\end{equation}
where the detected direction is denoted as $\textit{\textbf{n}}=\sin\theta\cos\varphi\textit{\textbf{i}}+\sin\theta\sin\varphi\textit{\textbf{j}}+\cos\theta\textit{\textbf{k}}$. Note that the  normalized $\omega=\omega/\omega_{0}$ is used for the radiation frequency.

By a careful examination for the electron motion with periodic part and drift displacement part, it is not difficult to find that there is a dimensionless fundamental frequency $\omega_{1}=2\pi/(T-\textit{\textbf{n}} \cdot \textit{\textbf{r}}_{0}+z_{0})$ in the emission spectra, where the periodic part of the electron's motion in the combing field has a period as $T=2n\pi$ associated to the magnetic resonance parameter $n$. Therefore, we can simplify Eq.(\ref{eq15}) by an infinite series of delta functions to calculate the emission power along any direction \cite{PRA-94-052102}. Accordingly the dimensionless vector $\textit{\textbf{F}}(\omega)$ can be expanded at the orders of the harmonics of $\omega_{1}$ \cite{POP-9-4325} as
\begin{equation}
\label{eq16}\textit{\textbf{F}}(\omega)=\frac{1}{\varsigma}\sum_{m=-\infty}^{+\infty}\textit{\textbf{F}}_{m}\delta\left (\omega-m\omega_{1} \right ),
\end{equation}
where the $m^{\rm{th}}$ amplitude is
\begin{equation}
\label{eq17}\textit{\textbf{F}}_{m}=\omega_{1}\int_{\eta_{in}}^{\eta_{in}+T}d\eta\textit{\textbf{p}}(\eta)\exp{\left \{ im2\pi h(\eta) \right \}},
\end{equation}
with a function of
\begin{equation}
\label{eq18}h(\eta)=\frac{\eta-\textit{\textbf{n}}\cdot\textit{\textbf{r}}(\eta)+z(\eta)}{T-\textit{\textbf{n}}\cdot\textit{\textbf{r}}_{0}+z_{0}},
\end{equation}
where $\textit{\textbf{r}}_{0}=(0,0,z_{0})=\left(0,0,T\left[\frac{n^{2} a^{2}}{2 \varsigma^{2}}\left(\varepsilon_{1}^{2}\left(\sin ^{2} \eta_{i n}+\frac{1}{2}\right)+\varepsilon_{2}^{2}\left(\cos ^{2} \eta_{i n}+\frac{1}{2}\right)\right)+\left(\frac{1}{2 \varsigma^{2}}-\frac{1}{2}\right)\right]\right)$ is the drift displacement vector of the electron during one period.

Finally, we can obtain the radiation spectra in units of Gaussian, erg/s per unit solid angle, by integrating the frequency $\omega$, in particular, the $m^{\rm{th}}$ harmonic is
\begin{equation}
\label{eq19}\frac{d^2 I_{m} }{d\Omega dt }=\frac{e^2\omega_{0}^{2}}{4\pi ^2c}\frac{1}{\varsigma^{2}}\left(m\omega_1\right)^2\left|\textit{\textbf{n}}\times \left[\textit{\textbf{n}}\times\textit{\textbf{F}}_m \right ]\right|^2,
\end{equation}
where the last term can be written further as
\begin{equation}
\label{eq20}\left|\textit{\textbf{n}}\times \left[\textit{\textbf{n}}\times\textit{\textbf{F}}_m \right ]\right|^2=\left|\textit{\textbf{n}} \left[\textit{\textbf{n}}\cdot\textit{\textbf{F}}_m \left(F_{mx},F_{my},F_{mz}\right)\right] -\textit{\textbf{F}}_m \left(F_{mx},F_{my},F_{mz}\right)\right|^2.
\end{equation}
Thus, based on Eq.(\ref{eq8})-Eq.(\ref{eq20}), the angular distributions of NTS can be calculated numerically when an electron moves in the combining field. For simplicity, appropriate parameters are chosen to make the radiation reaction effect (RRE) negligible and the radiated power per unit of solid angle is normalized by $\textit{e}^{2}\omega_{0}^{2}/4\pi^{2}\textit{c}$.

\begin{figure}[htbp]\suppressfloats
\includegraphics[width=10cm]{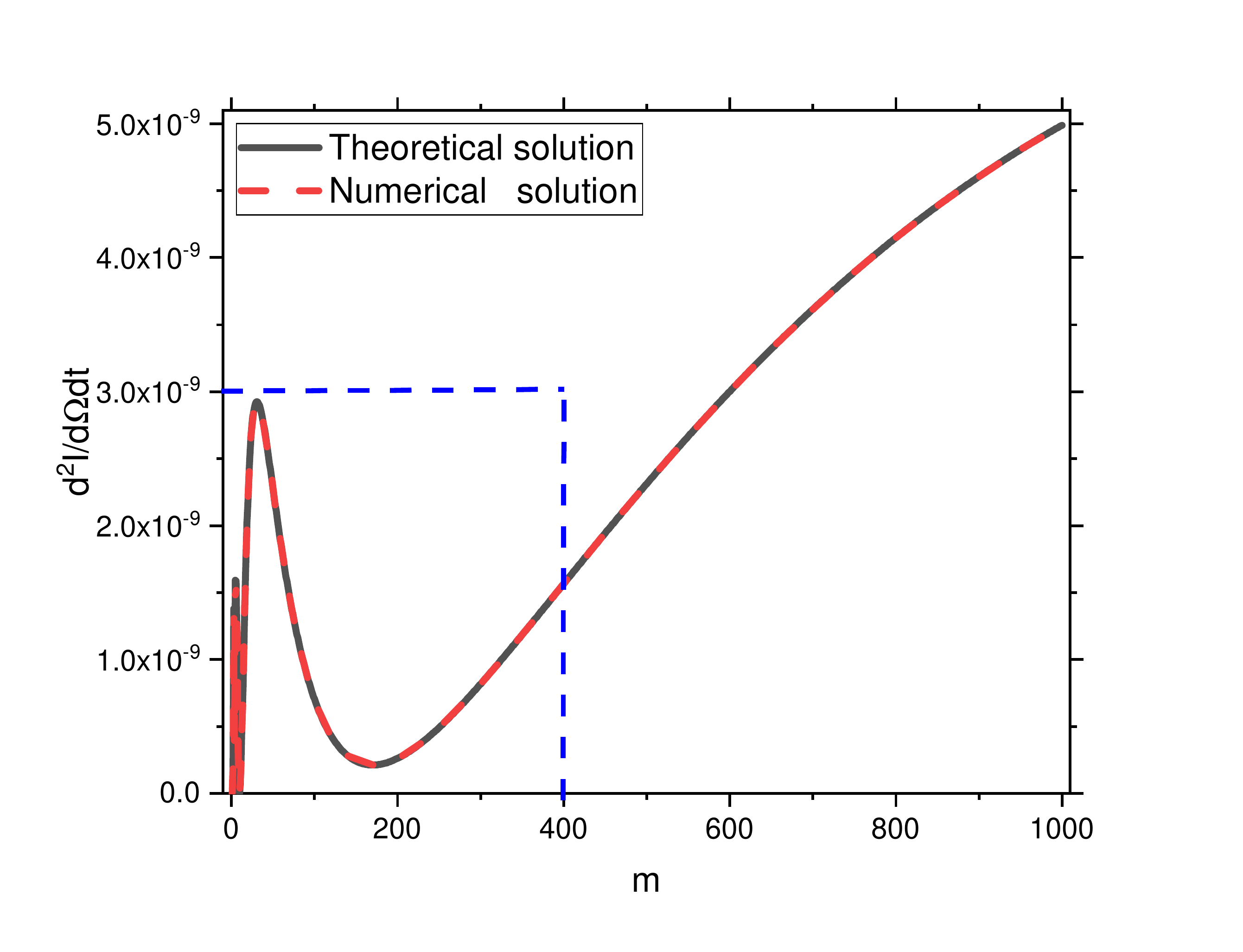}
\caption{\label{Fig2}(colour online)Compare the theoretical solution with the numerical solution : The backscattering spectra of the $m^{\rm{th}}$ harmonic when $\alpha=1$, $n=5$, $a=2$, $p_{z0}=1$, $\eta_{in}=0$.}
\end{figure}

Before giving a detailed study and description of our numerical results and analysis, we should emphasize that there is a delicate distinguish choosing for the laser field intensity in the following two aspects about the dependence of NTS on $\alpha$ in Sec.III and the angular distributions when $\alpha=0.5$ is fixed in Sec.IV. We know that the laser intensity $I_{0}=[{1.38 a^{2}(1+\alpha^2)}/{\lambda[{\mathrm{\mu m}}]^{2}}]\times10^{18}\mathrm{W/cm^{2}}$, where $a=eA_{0}/(\sqrt{1+\alpha^2}mc^{2})=a_{0}/\sqrt{1+\alpha^2}$ is the normalized vector potential amplitude and $\lambda=1\mathrm{\mu m}$ is the wavelength of the laser field which is a constant in our study. Obviously, only $a$ changes with $\alpha$. So we use $I_{0}$ in Sec.III and $a$ in Sec.IV.

In general, the final spectra of NTS via explicit analytical expression in combining field with a general elliptical polarization is hard to get. However, we can use Romberg integral algorithm to calculate $d^2 I_{m} /d\Omega dt$ and get the spatial distribution features of the NTS numerically.

Before starting our study for elliptical laser field, in order to check our validity of algorithm, we use Romberg algorithm to calculate the radiation spectrum of Thomson backscattering in combining field with circularly polarized laser. The comparison of theoretical and numerical solutions is shown in Fig.\ref{Fig2}, we can clearly see that they are in perfect agreement. From the Fig.\ref{Fig2}, we have checked that the numerical solution of harmonic order $1\sim400$ inside the blue border is consistent with the theoretical solution in Fig.3(d) of the Ref.\cite{PRA-94-052102}. Moreover, by the way, we correct a trivial typo of Ref.\cite{PRA-94-052102}, which should be written as $d^2I_{m}/d\Omega dt=\frac{e^2\omega_{0}^{2}}{4\pi^2c}\frac{2}{\varsigma^{2}}\left(m\omega_1\right)^2\left(\pi \omega_1n^2a\right)^2\left[A^2_{(m,n)}+B^2_{(m,n)}\right]$, where the typo occurs for the power index of $m$ as $m^4$. Note that this typo has not affect the numerical results of Ref.\cite{PRA-94-052102} since the formula in the routine is correct.

\section{Dependence of NTS on ellipticity $\alpha$}

In this section, we investigate the dependence of the electron trajectory, the fundamental frequency, the maximum radiation power in spatial distribution and corresponding spatial angle on $\alpha$.

\begin{figure}[htbp]\suppressfloats
\includegraphics[width=7.5cm,height=6cm]{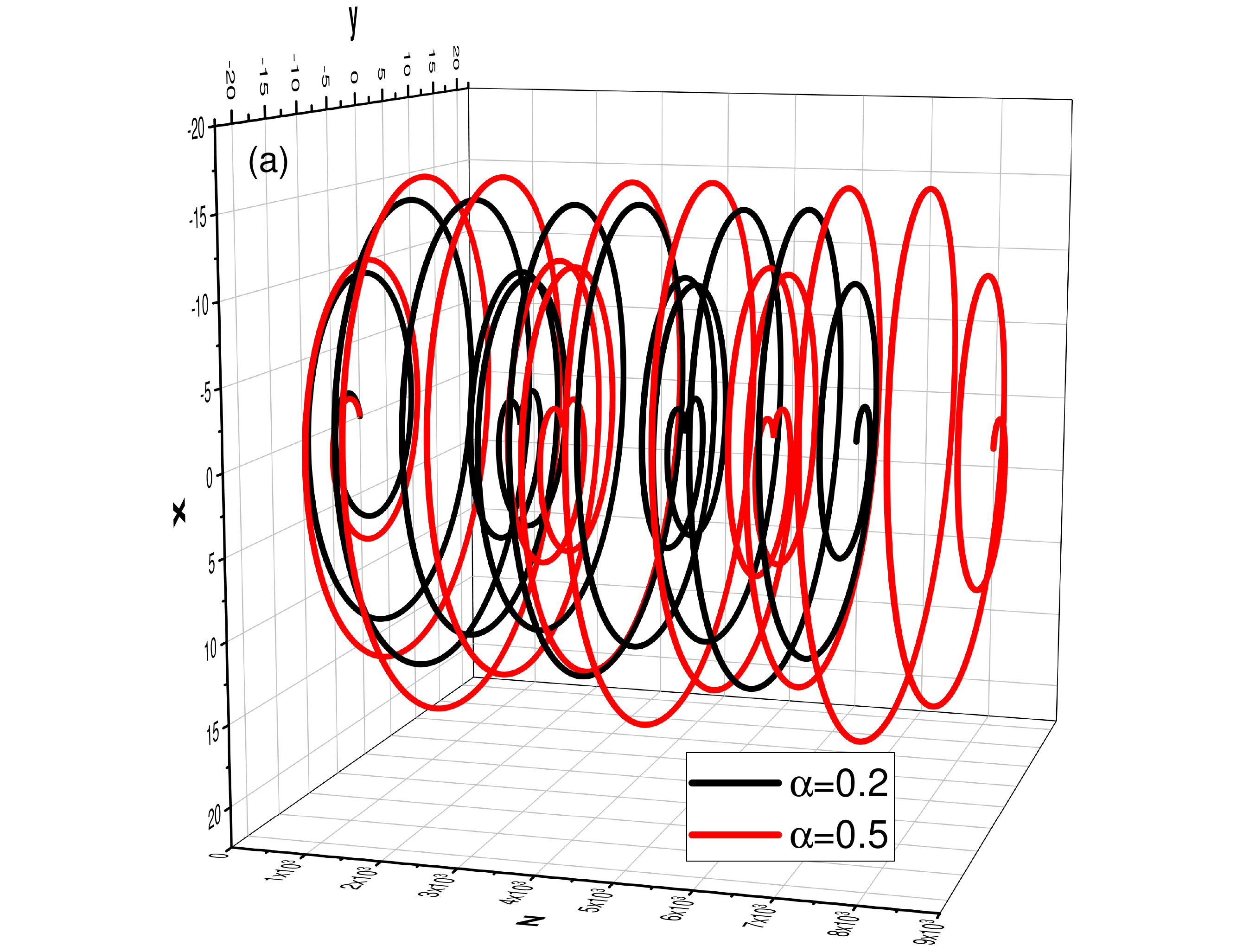}
\includegraphics[width=7.5cm,height=6cm]{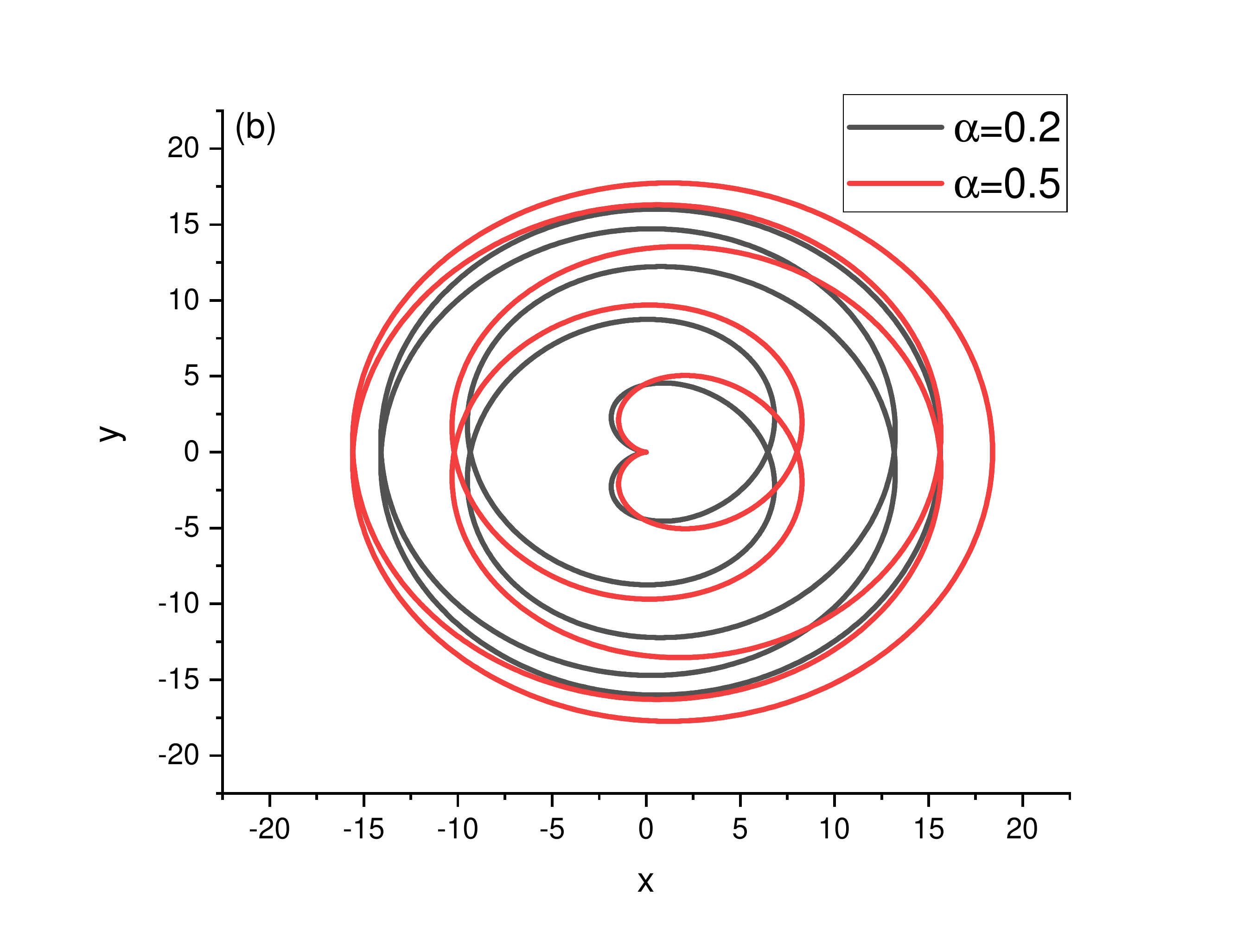}
\includegraphics[width=7.5cm,height=6cm]{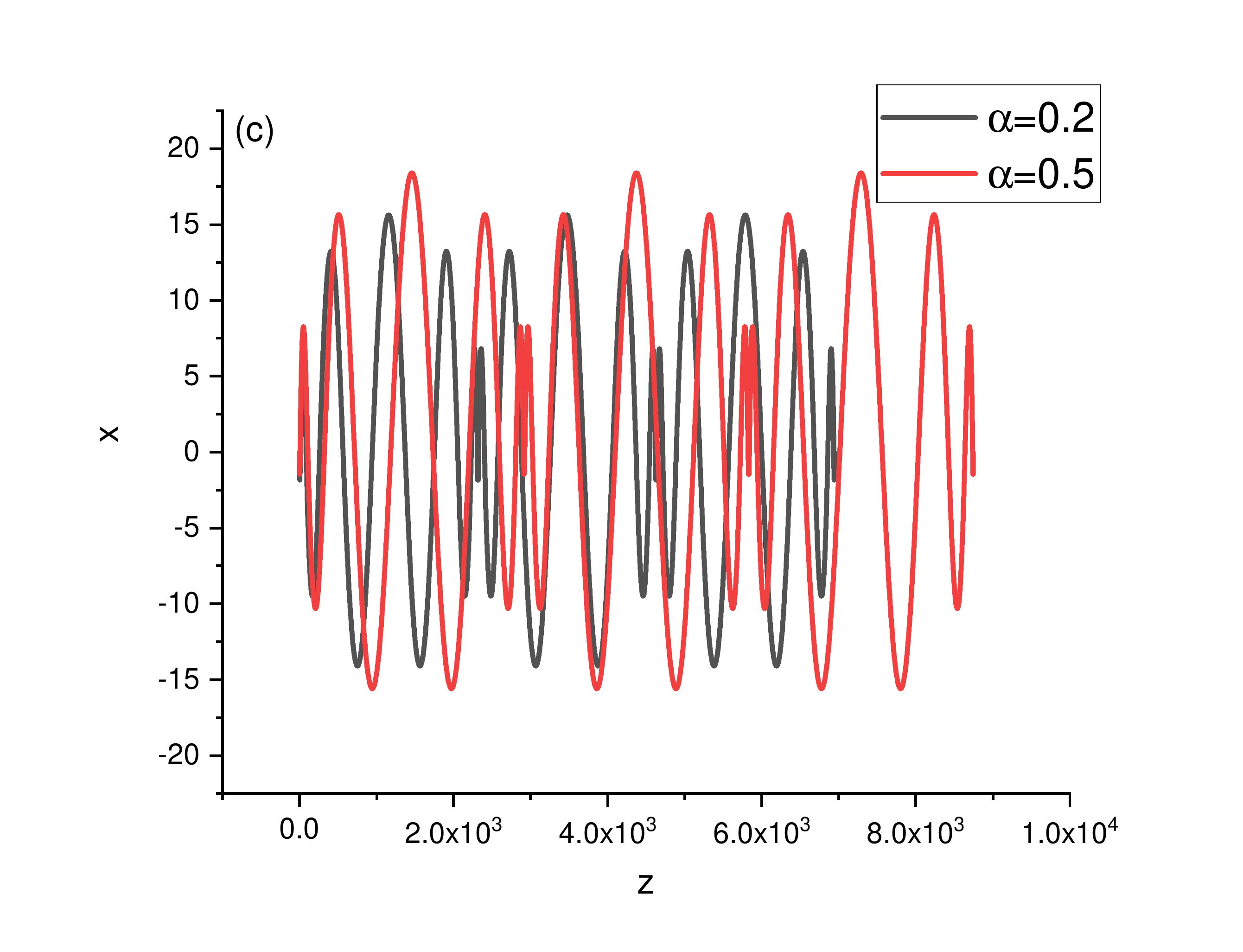}
\includegraphics[width=7.5cm,height=6cm]{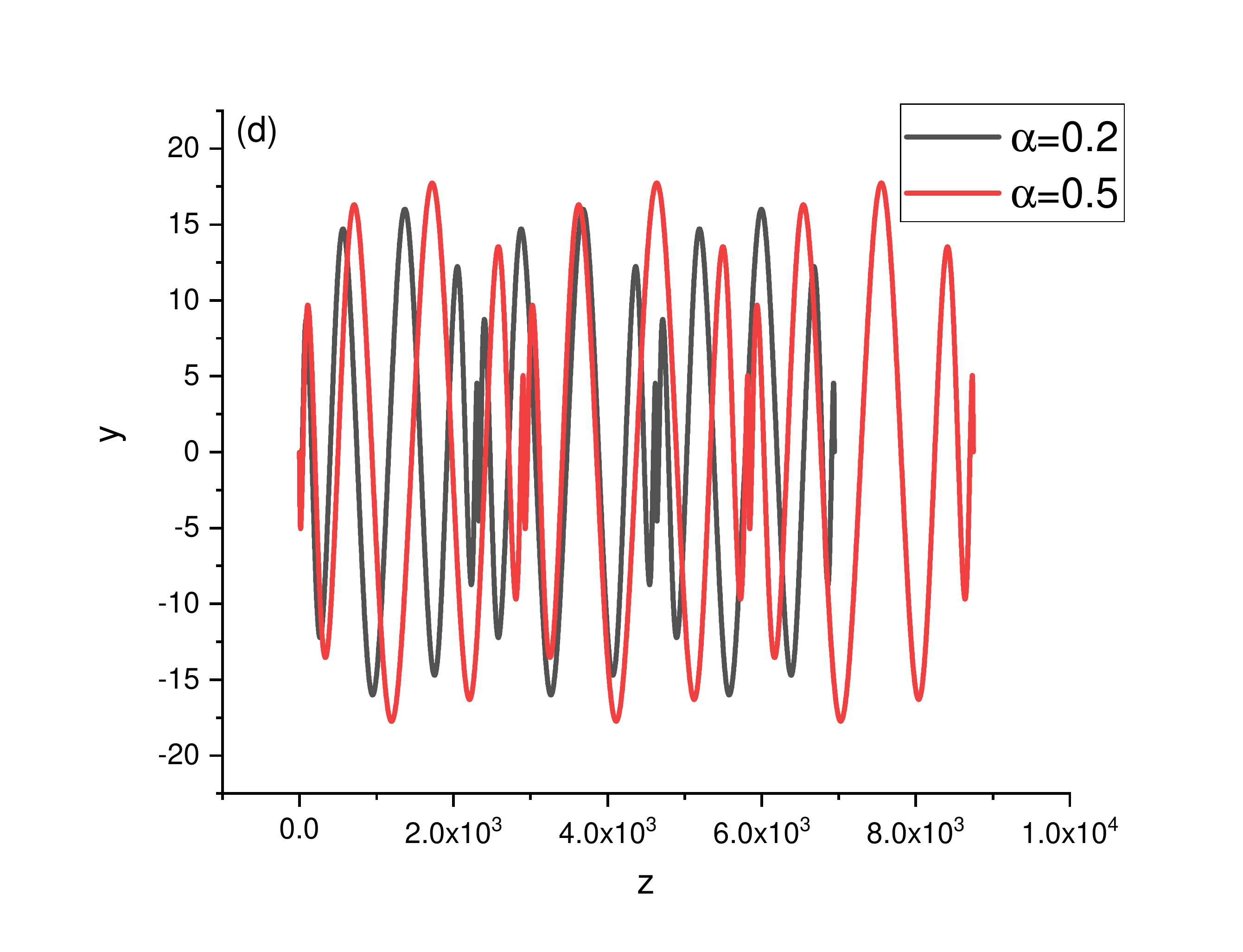}
\caption{\label{Fig3}(colour online) Electron's trajectories (a)-(d) in the combining laser and magnetic fields when $\alpha=0.2,0.5$. The initial axial momentum $p_{z0}=0$, the laser intensity as $I_{0}=1.10\times10^{19}\mathrm{W/cm^{2}}$ and the resonance parameter $n=5$. The initial phase $\eta_{in}=0$, The phase $\eta$ varies from 0 to $30\pi$.}
\end{figure}

\subsection{The electron trajectories}

First of all, the electron trajectory in the combining field with an elliptically polarization laser and a background magnetic field is studied. As an example, the electron trajectories for various $\alpha$ are shown in Fig.\ref{Fig3}, where the initial axial momentum $p_{z0}$, the laser intensity $I_{0}$, the resonance parameter $n$ and the initial phase $\eta_{in}$ are set as $p_{z0}=0,I_{0}=1.10\times10^{19}\mathrm{W/cm^{2}},n=5$ and $\eta_{in}=0$, respectively. Due to the periodicity of phase $\eta$, the electron trajectories are calculated over three optical cycles(the phase $\eta$ varies from 0 to $30\pi$). Notably, the following results have nothing to do with the optical cycles. The three dimensional trajectories and the projection to the $xy$ plane of the electron are shown in Fig.\ref{Fig3}(a) and 3(b) respectively. Obviously, the electron moves periodically in a spiral shape in the combining field. As shown in Fig.\ref{Fig3}(c) and 3(d), the projections of the electron trajectories in both the $xz$ and $yz$ planes are oscillating and the trajectories drift in $z$ direction. It can be seen that over one period $\textit{T}$, the electron undergoes a net displacement $\boldsymbol{r}_{0}=(0,0,z_{0})$, which is consistent with our theory mentioned in Sec.II. By comparison, we find that the electron moves like drawing circles around and the shapes of the trajectories remain unchanged for different $\alpha$. But it can be seen that the electron trajectory drift displacement and maximum oscillation amplitude when $\alpha=0.5$ are greater than that when $\alpha=0.2$.

\begin{figure}[htbp]\suppressfloats
\includegraphics[width=15cm]{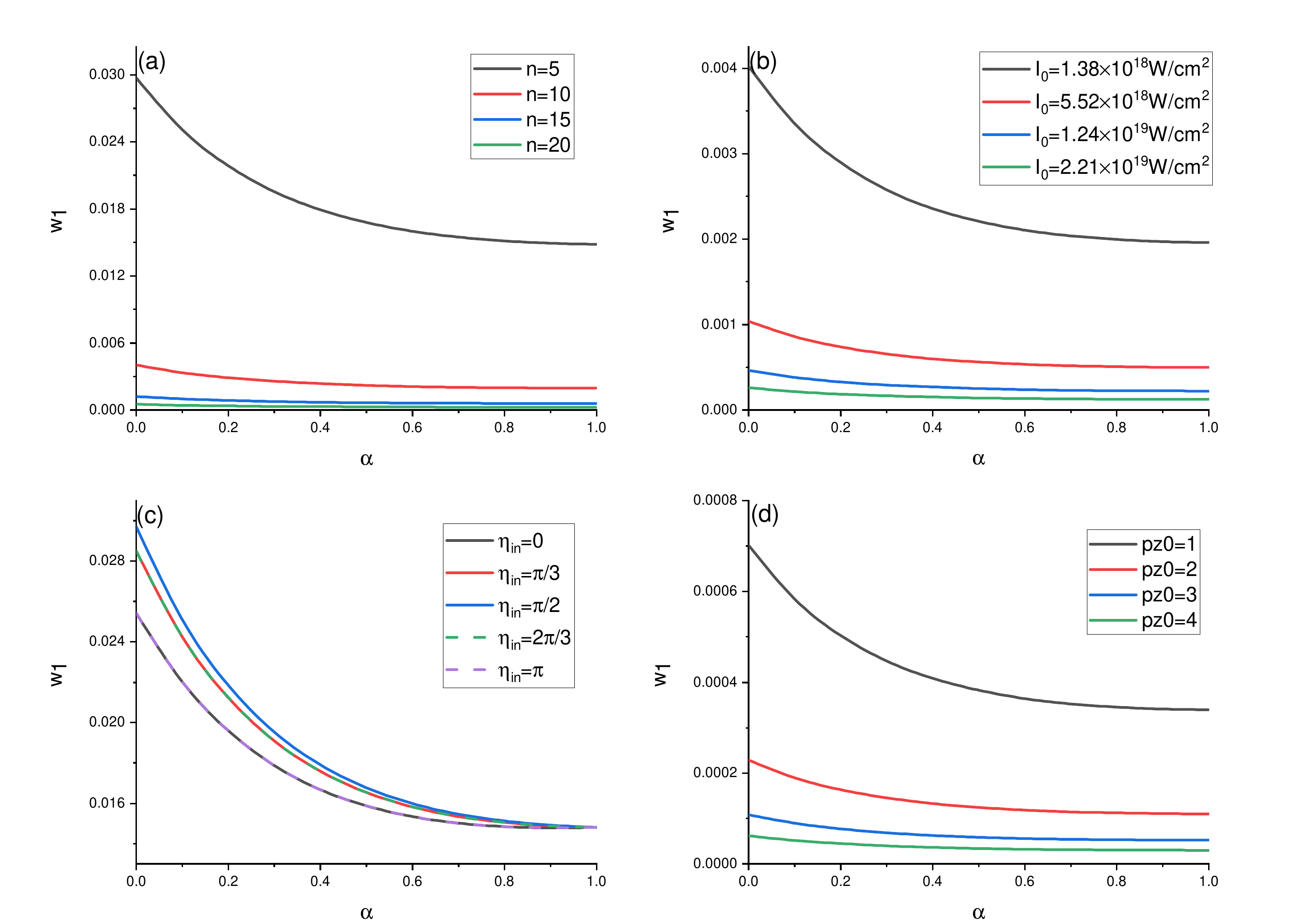}
\caption{\label{Fig4}(colour online) The relation of the fundamental frequency (normalized by $\omega_{0}$) to ellipticity $\alpha$ with different resonance parameter $n$, laser intensity $I_{0}$, initial axial momentum $p_{z0}$, and initial phase $\eta_{in}$. The other parameters are (a) $I_{0}=1.38\times10^{18}\mathrm{W/cm^{2}}$, $p_{z0}=0$, $\eta_{in}=\pi/2$; (b) $n=10$, $p_{z0}=0$, $\eta_{in}=\pi/2$; (c) $n=5$, $I_{0}=1.38\times10^{18}\mathrm{W/cm^{2}}$, $p_{z0}=0$; (d) $n=10$, $I_{0}=1.38\times10^{18}\mathrm{W/cm^{2}}$, $\eta_{in}=\pi/2$.}
\end{figure}

\subsection{The fundamental frequency when $\theta=\pi/2$}

Secondly, the dependence of fundamental frequency $\omega_{1}$ to $\alpha$ with variations in the magnetic resonance parameter $n$, laser intensity $I_{0}$, initial phase $\eta_{in}$ and initial axial momentum $p_{z0}$ are shown in Fig.\ref{Fig4}(a),(b),(c) and (d), respectively, where $\alpha$ varies from $0$ to $1$ and $\theta=\pi/2$. It can be seen that $\omega_{1}$ decreases with the increase of $\alpha$. Meanwhile, the $\omega_{1}$ of linearly polarized laser ($\alpha=0$) is the largest and obviously dependent on $\alpha$. And $\omega_{1}$ decreases with the increase of $n$, $I_{0}$ and $p_{z0}$ when other parameters are fixed, and at the same time, the dependence of $\omega_{1}$ on $\alpha$ decreases. Fig.\ref{Fig4}(d) shows that $\omega_{1}$ increases monotonically with the increase of $\eta_{in}$ in the range $[0,\pi/2]$ and decreases in the range $[\pi/2,\pi]$. So, $\omega_{1}$ changes periodically with the change of the $\eta_{in}$ and the period is $\pi$. In addition, the red solid line when $\eta_{in}=\pi/3$ coincides exactly with the green dotted line when $\eta_{in}=2\pi/3$ and the black solid line when $\eta_{in}=0$ coincides exactly with the purple dotted line when $\eta_{in}=\pi$, which means that $\omega_{1}$ vs $\eta_{in}$ is symmetrical about $\eta_{in}=\pi/2$ and the peak  of $\omega_{1}$ is achieved when $\eta_{in}=\pi/2$. It is worth noting that $\omega_{1}$ is almost independent of the $\eta_{in}$ when $\alpha$ is greater than $0.8$, and it is the same for circularly polarized laser ($\alpha=1$). On the other hand, obviously, the largest fundamental frequency is achieved for the linear polarized lase ($\alpha=0$) when $\eta_{in}=\pi/2$. The conclusion above is consistent with that by Jiang \textit{et al.} \cite{EPL-117-44002}. Beside the wide studied situation for Thomson backscattering ($\theta=\pi$), it is noted that the $\omega_{1}$ increases with the decrease of $\theta$ which is an obvious by its theoretical expression.

\begin{figure}[htbp]\suppressfloats
\includegraphics[width=15cm]{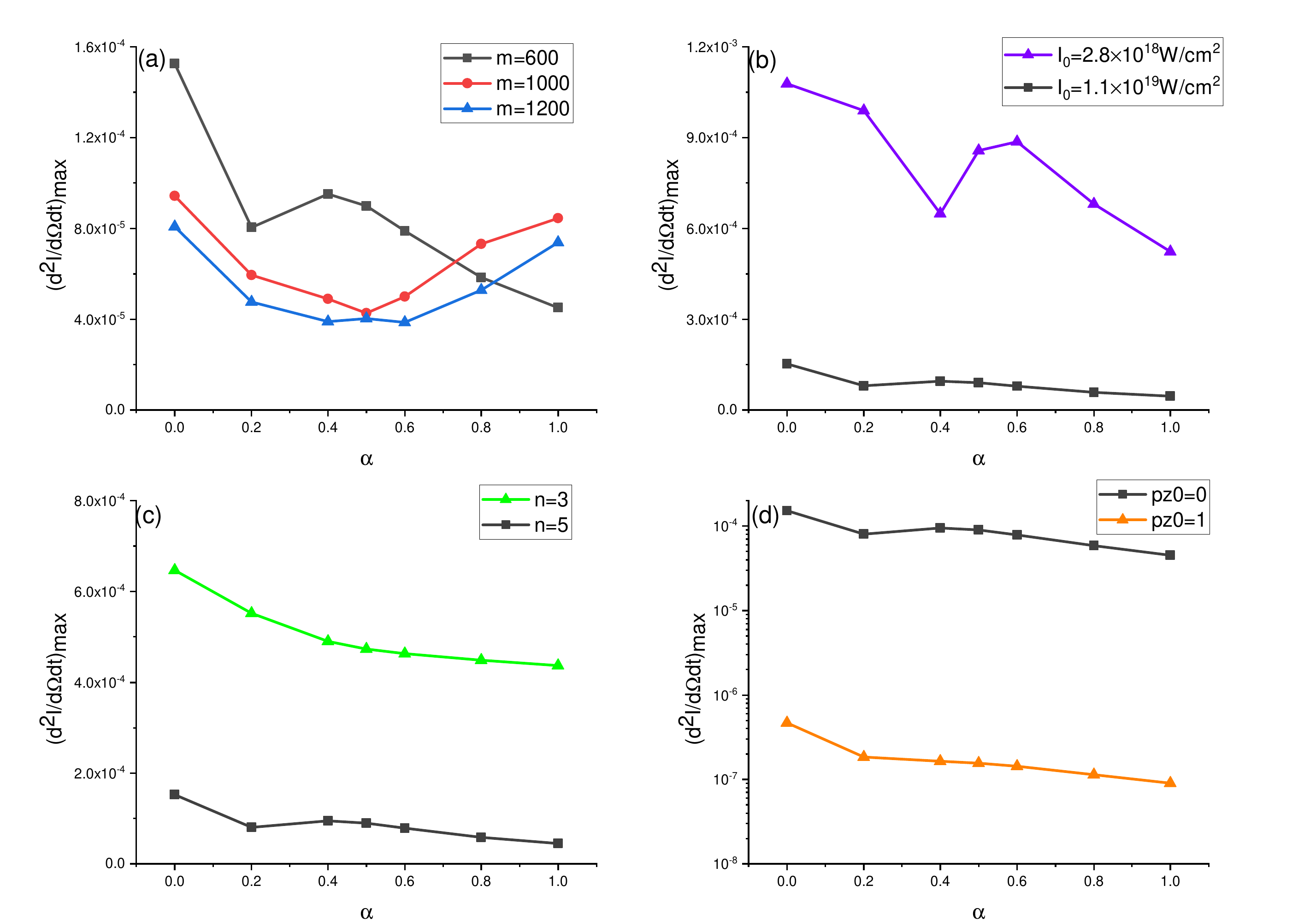}
\caption{\label{Fig5}(colour online) The relation of the maximum radiated power in the $xy$ plane to ellipticity $\alpha$ with different harmonic order number $m$, laser intensity $I_{0}$, resonance parameter $n$ and initial axial momentum $p_{z0}$ when the initial phase $\eta_{in}=0$ is fixed. The other parameters are (a) $I_{0}=1.10\times10^{19}\mathrm{W/cm^{2}}$, $n=5$, $p_{z0}=0$; (b) $m=600$, $n=5$, $p_{z0}=0$; (c) $m=600$, $I_{0}=1.10\times10^{19}\mathrm{W/cm^{2}}$, $p_{z0}=0$; (d) $m=600$, $I_{0}=1.10\times10^{19}\mathrm{W/cm^{2}}$, $n=5$.}
\end{figure}

\begin{figure}[htbp]\suppressfloats
\includegraphics[width=15cm]{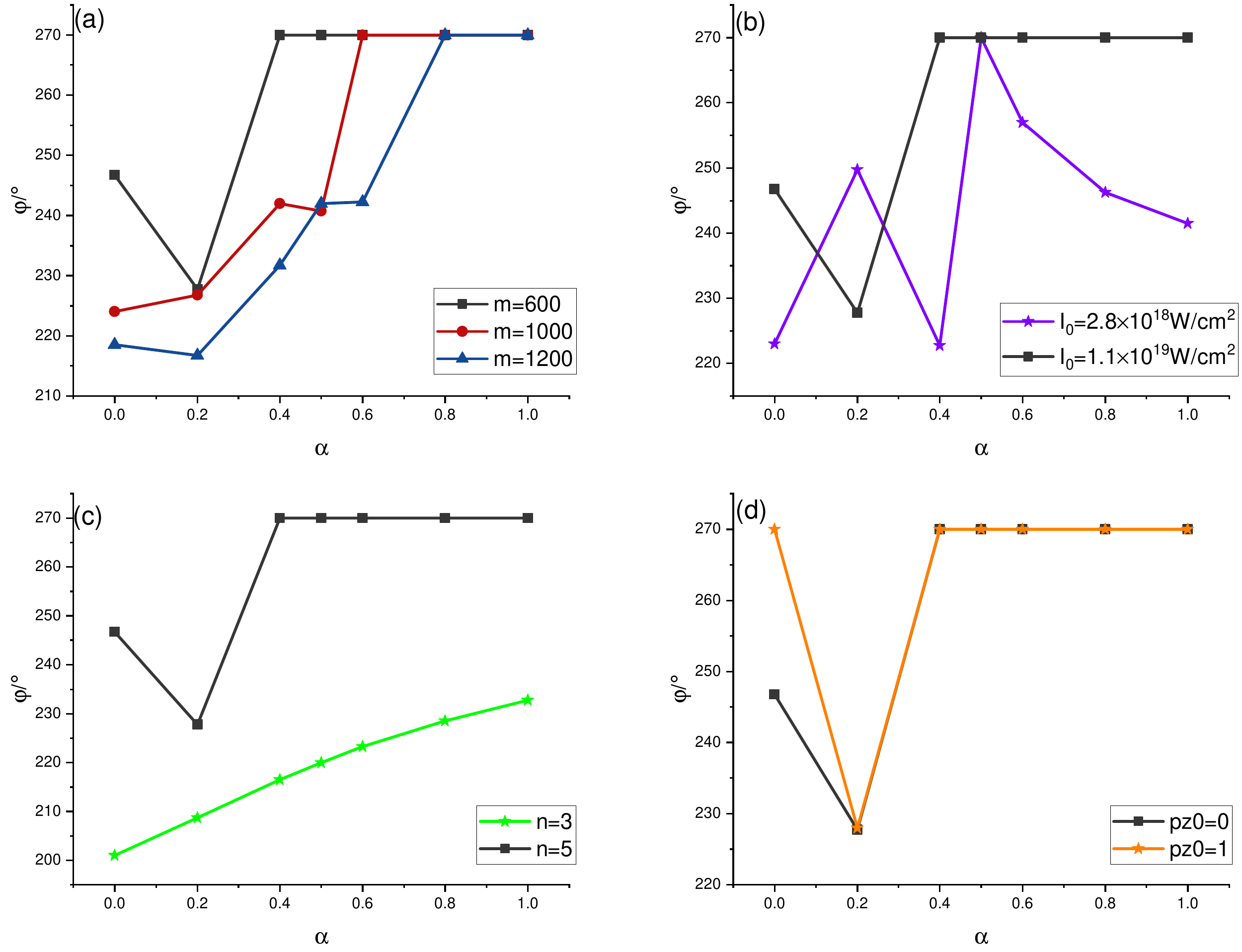}
\caption{\label{Fig6}(colour online) The relation of the azimuthal angle $\varphi$ according to the maximum radiated power in the $xy$ plane to ellipticity $\alpha$ with different harmonic order number $m$, laser intensity $I_{0}$, resonance parameter $n$ and initial axial momentum $p_{z0}$. Here the initial phase $\eta_{in}=0$. (a) $I_{0}=1.10\times10^{19}\mathrm{W/cm^{2}}$, $n=5$, $p_{z0}=0$; (b) $m=600$, $n=5$, $p_{z0}=0$; (c) $m=600$, $I_{0}=1.10\times10^{19}\mathrm{W/cm^{2}}$, $p_{z0}=0$; (d) $m=600$, $I_{0}=1.10\times10^{19}\mathrm{W/cm^{2}}$, $n=5$.}
\end{figure}

\subsection{The maximum radiation power in spatial distribution and corresponding spatial angle}

Thirdly, the maximum radiation power in spatial distribution and corresponding spatial angle with different harmonic order number $m$, laser intensity $I_{0}$, resonance parameter $n$, and initial axial momentum $p_{z0}$ are investigated. It is mainly divided into two parts, one is the relation of the maximum radiated power and corresponding azimuthal angle $\varphi$ in the $xy$ plane to $\alpha$ when polar angle $\theta=\pi/2$, the other is the relation of the maximum radiated power and corresponding $\theta$ to $\alpha$ when $\varphi=0$. By the way, for the more results about the angular distribution one can refer to \cite{supplement}.

As can be seen from Fig.\ref{Fig5}, the maximum radiated power in the $xy$ plane changes nonlinearly with the increase of $\alpha$. Clearly, the value of $(d^{2}I/d\Omega dt)_{max}$ decreases with the increase of $\alpha$, and it oscillates when $\alpha=0.2$-$0.6$. In addition, the $(d^{2}I/d\Omega dt)_{max}$ decreases and then increases when $m=1000$ or/and $m=1200$, which reflects the characteristics of nonlinear change.
In the direction perpendicular to the propagation of the laser, the maximum radiation power of the linearly polarized laser is higher than that of the elliptically and circularly ones. Moreover, the elliptically polarized laser is better than the circularly polarized laser. The value of $(d^{2}I/d\Omega dt)_{max}$ has little difference in order of magnitude under different $\alpha$. When other parameters are fixed, the $(d^{2}I/d\Omega dt)_{max}$ decreases with the increase of $m$, $n$, $I_{0}$ and $p_{z0}$. However, the $(d^{2}I/d\Omega dt)_{max}$ decreases and then increases when $\alpha=0.8$ or/and $\alpha=1.0$, as in Fig.\ref{Fig5}(a). The oscillation of the maximum radiation power with the change of $\alpha$ is more obvious when $I_{0}=2.76\times10^{18}[\mathrm{W/cm^{2}}]$ (Fig.\ref{Fig5}(b)). More details when $\alpha=0.5$ will be discussed in the following Sec.IV.

\begin{figure}[htbp]\suppressfloats
\includegraphics[width=15cm]{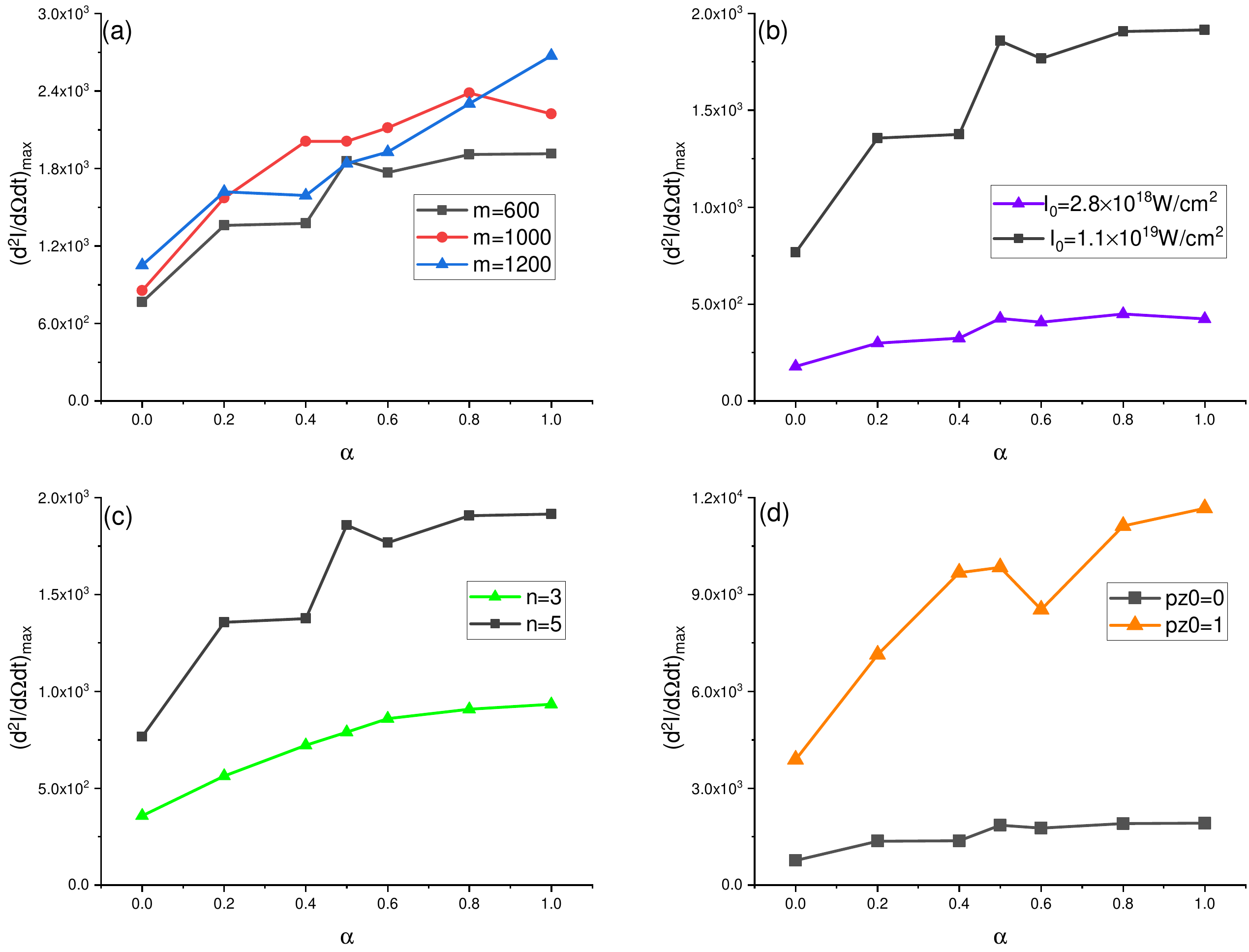}
\caption{\label{Fig7}(colour online) The relation of the maximum radiated power when $\varphi=0$ to ellipticity $\alpha$ with different harmonic order number $m$, laser intensity $I_{0}$, resonance parameter $n$ and initial axial momentum $p_{z0}$. Here the initial phase $\eta_{in}=0$. (a) $I_{0}=1.10\times10^{19}\mathrm{W/cm^{2}}$, $n=5$, $p_{z0}=0$; (b) $m=600$, $n=5$, $p_{z0}=0$; (c) $m=600$, $I_{0}=1.10\times10^{19}\mathrm{W/cm^{2}}$, $p_{z0}=0$; (d) $m=600$, $I_{0}=1.10\times10^{19}\mathrm{W/cm^{2}}$, $n=5$.}
\end{figure}

\begin{figure}[htbp]\suppressfloats
\includegraphics[width=15cm]{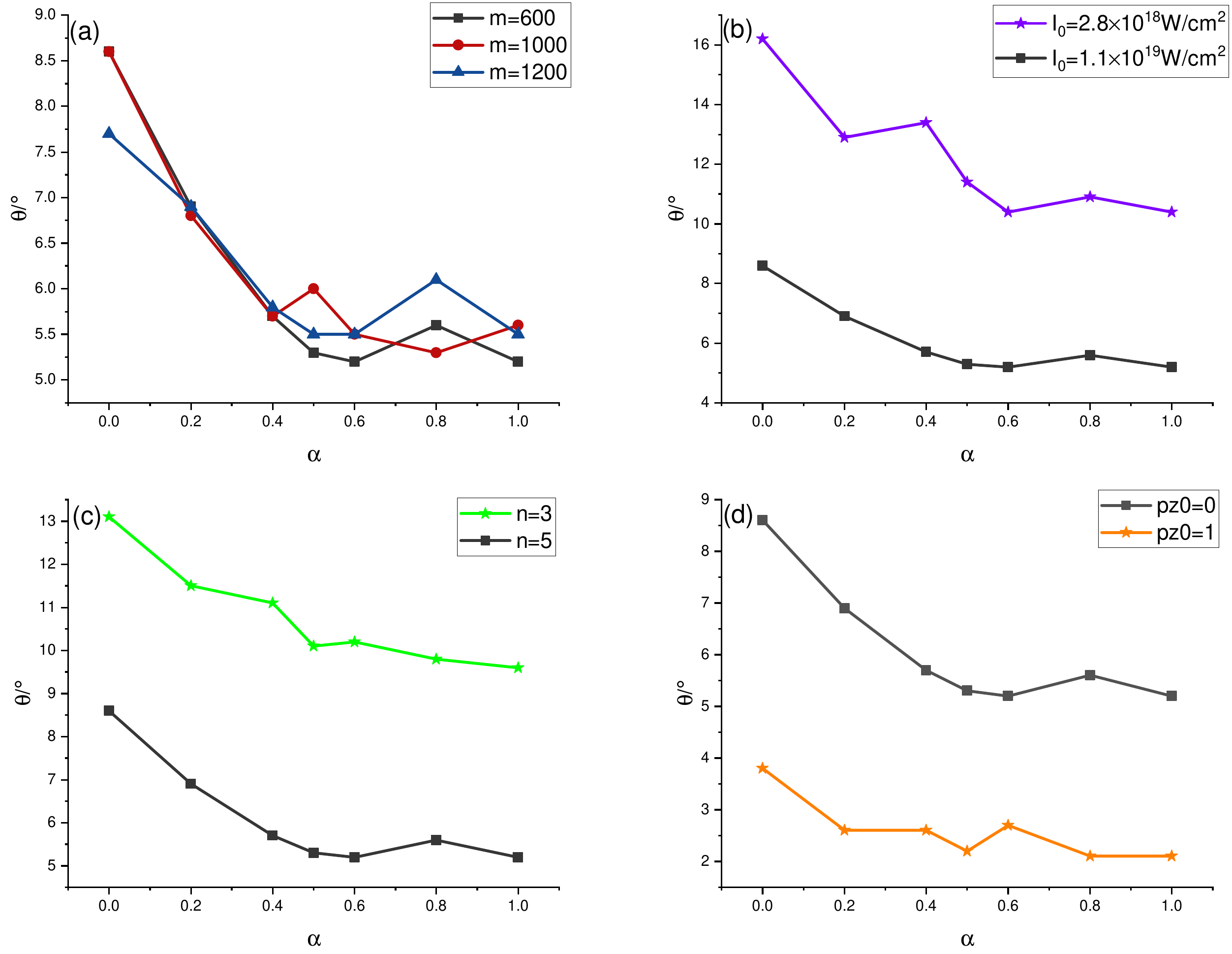}
\caption{\label{Fig8}(colour online) The relation of the polar angle $\theta$, corresponding to the maximum radiated power, when $\varphi=0$ to ellipticity $\alpha$ with different harmonic order number $m$, laser intensity $I_{0}$, resonance parameter $n$ and initial axial momentum $p_{z0}$. Here the initial phase $\eta_{in}=0$. (a) $I_{0}=1.10\times10^{19}\mathrm{W/cm^{2}}$, $n=5$, $p_{z0}=0$; (b) $m=600$, $n=5$, $p_{z0}=0$; (c) $m=600$, $I_{0}=1.10\times10^{19}\mathrm{W/cm^{2}}$, $p_{z0}=0$; (d) $m=600$, $I_{0}=1.10\times10^{19}\mathrm{W/cm^{2}}$, $n=5$.}
\end{figure}

The change of $\varphi$ according to the maximum radiated power in the $xy$ plane with $\alpha$ can be seen from Fig.\ref{Fig6}. As the radiation distribution in the $xy$ plane is symmetric about the $y$ axis \cite{supplement}, we only describe one of the angles in the range $180^{\circ}$ to $270^{\circ}$. The change of another angle with $\alpha$ is symmetric and will not be discussed below. The nonlinear change of $\varphi$ with $\alpha$ can be seen, and most of the $\varphi$ approach $270^{\circ}$ as $\alpha$ increases. However, it oscillates more obviously with the increase of $\alpha$ when $I_{0}=2.76\times10^{18}[\mathrm{W/cm^{2}}]$, and it changes linearly when $n=3$, as in Fig.\ref{Fig6}(c).

In Fig.\ref{Fig7}, the maximum radiated power when $\varphi=0$ increases with the increase of $\alpha$, and the pictures show the nonlinear characteristics of oscillation. It means that the linearly polarized laser in the $zx$ plane is the worst which is different from the $xy$ plane. However, we can see that the elliptically polarized laser is better than the circularly one for some parameter sets, for example, $\alpha=0.8$ when $m=1000$ shown in Fig.\ref{Fig7}(a), $\alpha=0.5,0.8$ when $I_{0}=2.76\times10^{18}[\mathrm{W/cm^{2}}]$ shown in Fig.\ref{Fig7}(b). With the increase of $m$, the higher order $(d^{2}I/d\Omega dt)_{max}$ of elliptically polarized laser decreases, the value of the maximum radiated power has little difference in order of magnitude under different $\alpha$. When other parameters are fixed, the $(d^{2}I/d\Omega dt)_{max}$ increases with the increase of $n$, $I_{0}$ and $p_{z0}$, which is also different from the $xy$ plane. From Fig.\ref{Fig8}, we can see that the $\theta$, corresponding to the maximum radiated power in the $zx$ plane, decreases with $\alpha$ increases. The pictures also show the nonlinear characteristics of oscillation, but the whole process is declining. With the increase of $m$, the value of the $(d^{2}I/d\Omega dt)_{max}$ has little difference under different $\alpha$. In addition, it is a descending process from $\alpha=0$ to $\alpha=0.4$, and oscillations are obvious at $\alpha=0.4$ to $\alpha=1.0$. As can be seen from Fig.\ref{Fig8}(b),(c),(d), the $\theta$ decreases with the increase of $n$, $I_{0}$ and $p_{z0}$, and it changes gently with the increase of $\alpha$ which is from $\alpha=0.5$ to $\alpha=1.0$.

\section{Angular distributions when ellipticity $\alpha=0.5$}

In this section we would examine in detail the angular distribution of the NTS when the middle ellipticity $\alpha=0.5$ is fixed. And through the numerical results, the maximum radiated power per unit of solid angle, the corresponding number of photons and photons brightness can be obtained in optimum emission  direction. By the way, for a convenience and simplicity, meanwhile without the loosing of the generality, besides $\alpha=0.5$, the initial phase $\eta_{in}=0$ for the electron is fixed in the following.

It is emphasized that the spatial distribution features of the NTS are researched from three perspectives. The first is the angular distributions with respect to $\varphi$, when $\theta=\pi/2$ is fixed. The second is the angular distributions with respect to $\theta$, when $\varphi=0$ is fixed. And thirdly, we look into the small angle spatial distribution by the contour plotting. It is very useful to write some important quantities in three cases like of, in the first case, the detected direction $\textit{\textbf{n}}=(\cos\varphi,\sin\varphi,0)$, $\omega_{1}={2\pi}/({2\pi n+z_{0}})$, and $h(\eta)=[{\eta-\cos\varphi x(\eta)-\sin\varphi y(\eta)}+z(\eta)]/[{2\pi n+z_{0}}]$. Similarly, in the second case, they are $\textit{\textbf{n}}=(\sin\theta,0,\cos\theta)$, $\omega_{1}={2\pi}/[{2\pi n+(1-\cos\theta)z_{0}}]$ and $h(\eta)=[{\eta-\sin\theta x(\eta)+(1-\cos\theta)z(\eta)}]/[{2\pi n+(1-\cos\theta)z_{0}}]$. In the third case, they are $\textit{\textbf{n}}=(\sin\theta\cos\varphi,\sin\theta\sin\varphi,\cos\theta)$,  $\omega_{1}={2\pi}/[{2\pi n+(1-\cos\theta)z_{0}}]$, and $h(\eta)=[{\eta-\sin\theta\cos\varphi x(\eta)-\sin\theta\sin\varphi y(\eta)+(1-\cos\theta)z(\eta)}]/[{2\pi n+(1-\cos\theta)z_{0}}]$. By the way, it is noted that the $\varphi$ and $\theta$ corresponding to the maximum radiated power would have different behavior in the third case from that in the former first and second cases.

\subsection{Angular distributions with respect to the azimuthal angle $\varphi$ when $\theta=\pi/2$}

\begin{figure}[htbp]\suppressfloats
\includegraphics[width=15cm]{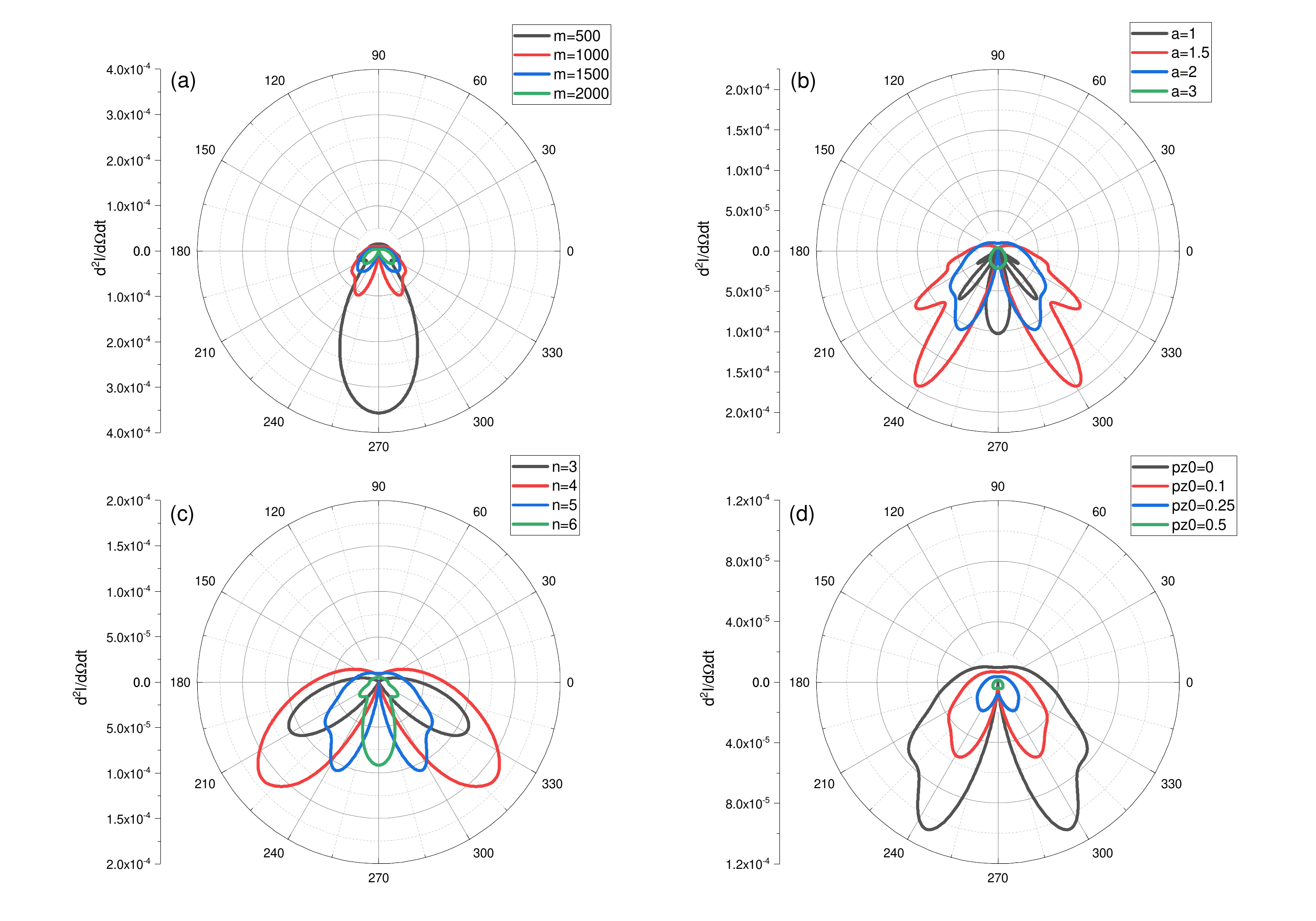}
\caption{\label{Fig9}(colour online) Spatial distributions of the NTS with respect to the azimuthal angle $\varphi$, for the different $m$, $a$, $n$ and $p_{z0}$, are shown in (a), (b), (c) and (d), respectively. The polar angle $\theta=\pi/2$. (a) $a=2$, $n=5$, $p_{z0}=0$; (b) $m=1000$, $n=5$, $p_{z0}=0$; (c) $m=1000$, $a=2$, $p_{z0}=0$; (d) $m=1000$, $a=2$, $n=5$.}
\end{figure}

In this section, the angular distributions of the emission power with respect to the azimuthal angle $\varphi$ when $\theta=\pi/2$ are investigated. The azimuthal angle distributions of the emitted power for various parameters are shown in Fig.\ref{Fig9}.

\begin{figure}[htbp]\suppressfloats
\includegraphics[width=15cm]{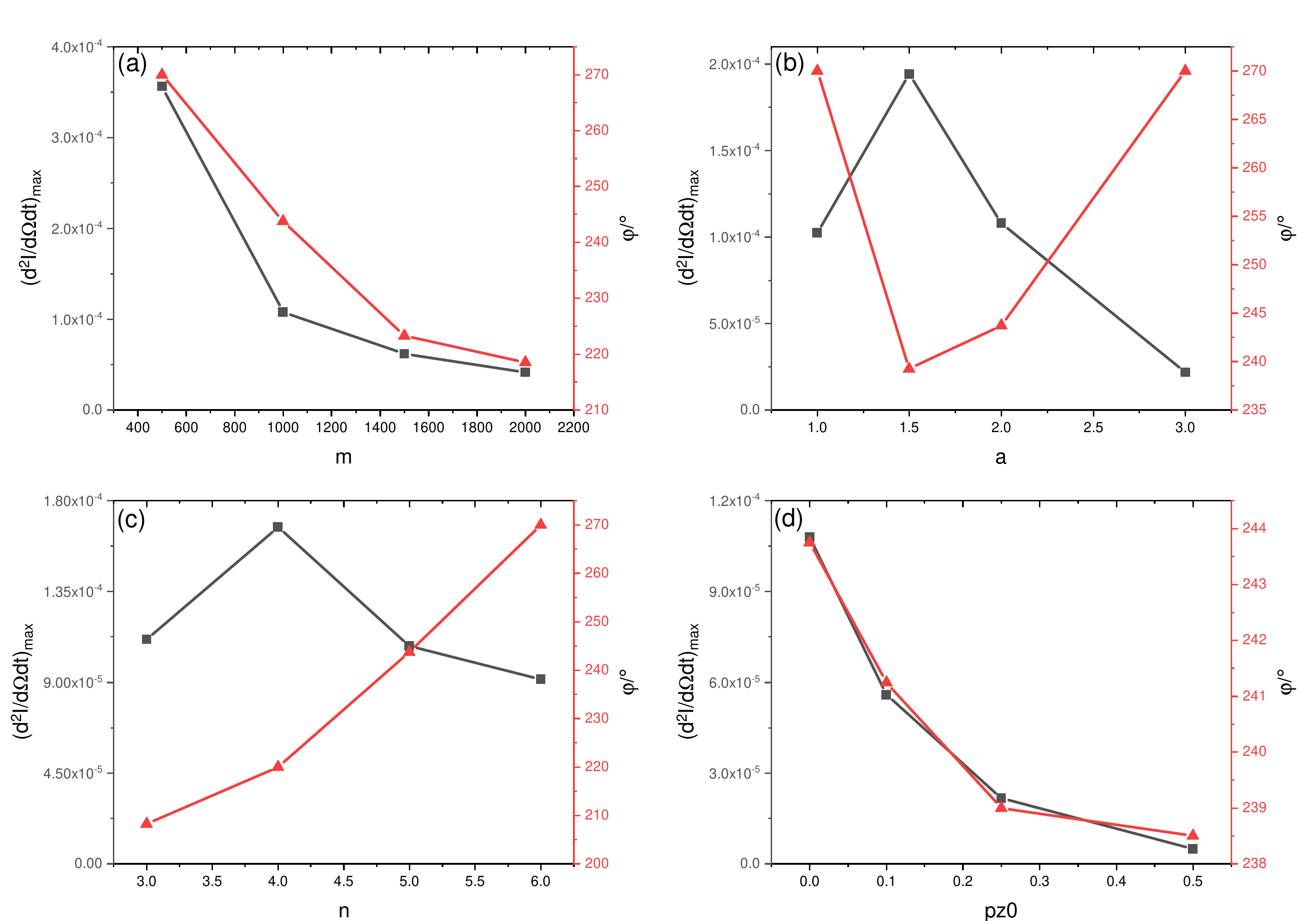}
\caption{\label{Fig10}(colour online) The maximum radiation power and corresponding $\varphi$ change with different $m$, $a$, $n$ and $p_{z0}$. The polar angle $\theta=\pi/2$. (a) $a=2$, $n=5$, $p_{z0}=0$; (b) $m=1000$, $n=5$, $p_{z0}=0$; (c) $m=1000$, $a=2$, $p_{z0}=0$; (d) $m=1000$, $a=2$, $n=5$.}
\end{figure}

From this figure, the different shapes of the spatial distribution of the radiation power are seen clearly. They are all symmetric with respect to $y$ axis whatever the $m$, $a$, $n$ and $p_{z0}$ are. Moreover, most of them show the BRP beside less of URP , while they have different shapes under different parameters. Fig.\ref{Fig9}(a) presents an unifoliate radiation pattern (URP) when $m=500$. In Fig.\ref{Fig9}(b), it oscillates significantly in the $xy$ plan when $a=1$, and it can turn into URP when $a=3$. Fig.\ref{Fig9}(c) shows the URP when $n=6$. In Fig.\ref{Fig9}(d), the shape is uniformly distributed in the $xy$ plane when $p_{z0}=0.5$, which is recovered to the results in Ref.\cite{Zhao-36-074101} as $p_{z0}$ increases. If the $p_{z0}$ is large, the radiation is approximately uniformly distributed with respect to $\varphi$. Meanwhile, the large radiation power mainly distributes between $\varphi=\pi$ and $\varphi=2\pi$. Now let us focus on the difference of the variation tendency of ${d^2 I}/{d\Omega dt }$ and ${d^2 I_{max} }/{d\Omega dt }$. Fig.\ref{Fig9}(a) and 9(d) show the angular distributions of the emission power with respect to $\varphi$ for different harmonic orders $m^{\rm{th}}$ and different $p_{z0}$, respectively. Their overall trend of ${d^2 I}/{d\Omega dt }$ decreases with the increase of $m$ or $p_{z0}$. Fig.\ref{Fig9}(b) and 9(c) show the angular distributions of the emission with respect to $\varphi$ for different $a$ and various $n$, respectively. Their overall trend of ${d^2 I}/{d\Omega dt }$ increases and then decreases with the increase of $a$ or $n$. They also show the nonlinear variation and the complexity of spatial radiation of the NTS.

In Fig.\ref{Fig10}, the maximum radiation power in spatial distribution and corresponding azimuthal angle with different $m$, $a$, $n$ and $p_{z0}$ are shown. As in Fig.\ref{Fig10}(a) and 10(d), ${d^2 I_{max} }/{d\Omega dt }$ decreases with the increase of $m$ or $p_{z0}$, and the corresponding $\varphi$ is smaller within a given range of parameters. In Fig.\ref{Fig10}(b) and 10(c), we can obviously see that their variation trend of ${d^2 I_{max}}/{d\Omega dt }$ increases and then decreases with the increase of $a$ or $n$. In Fig.\ref{Fig10}(b), the corresponding $\varphi$ of maximum radiation power varies in the range from $\varphi=240^{\circ}$ to $\varphi=270^{\circ}$. And in Fig.\ref{Fig10}(c), the $\varphi$ of maximum radiation power is approaching to the orientation of $\varphi=270^{\circ}$, which varies in the range from $\varphi=210^{\circ}$ to $\varphi=270^{\circ}$. It is worth noting that we can get the same maximum radiation power in two different directions due to the symmetry.

In general, the radiation angular distributions with respect to $\varphi$ shows the symmetry. The symmetry of radiation distributions will not be broken while the parameters change, the large radiation power mainly distributes between $\varphi=\pi$ and $\varphi=2\pi$. The study shows that the overall trend of ${d^2 I}/{d\Omega dt }$ increases and then decreases with the increase of $a$ or $n$ when $\alpha=0.5$. So we can obtain maximal radiation power and corresponding $\varphi$ by choosing suitable parameters in $xy$ plane which is perpendicular to the propagation direction of laser.

\begin{figure}[htbp]\suppressfloats
\includegraphics[width=15cm]{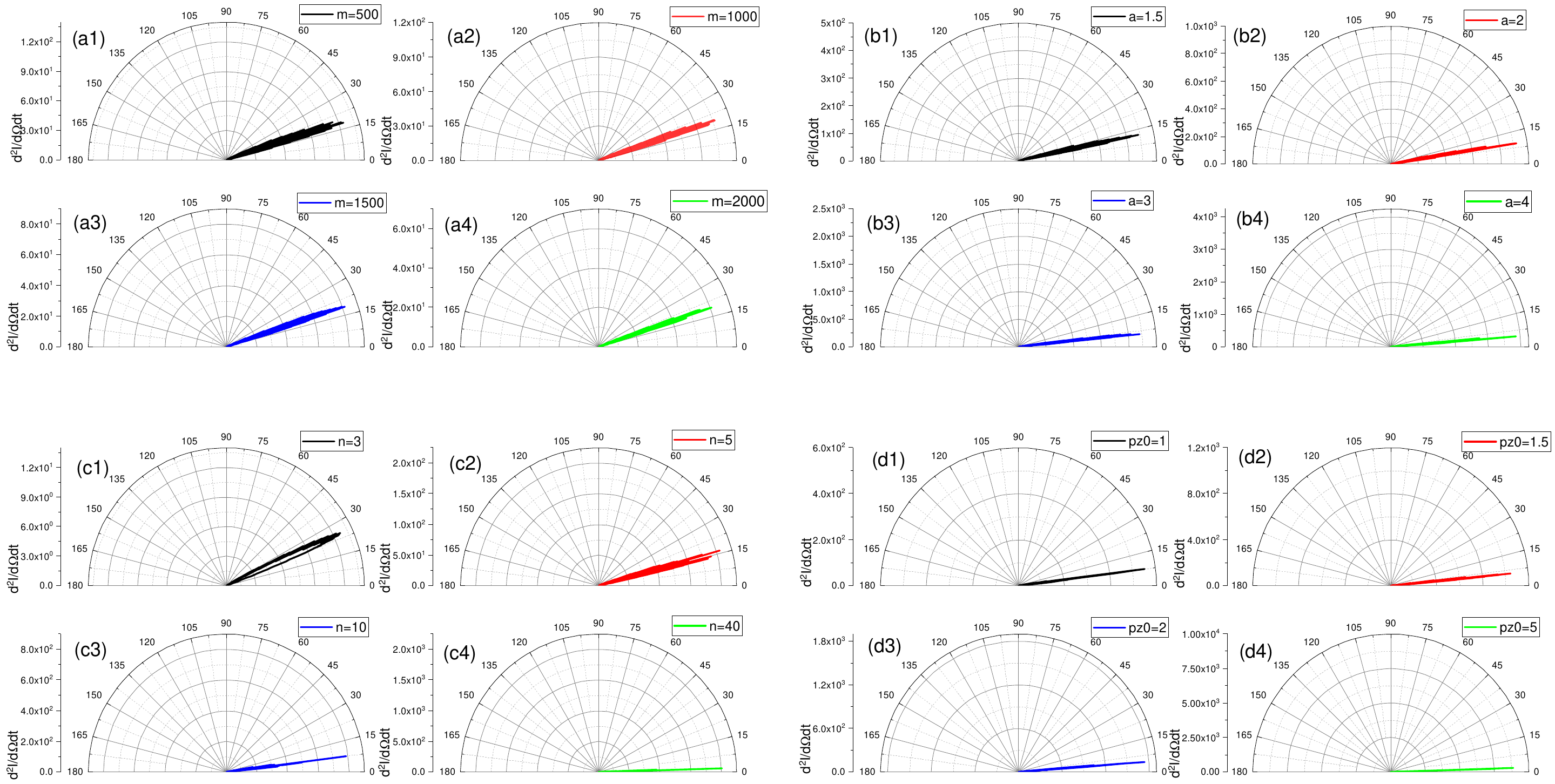}
\caption{\label{Fig11}(colour online) Spatial distributions of the NTS with respect to the polar angle $\theta$ ($\varphi=0$), for the different $m$, $a$, $n$ and $p_{z0}$, are shown in (a), (b), (c) and (d), respectively. (a) $a=1$, $n=4$, $p_{z0}=0$; (b) $m=1000$, $n=4$, $p_{z0}=0$; (c) $m=1000$, $a=1$, $p_{z0}=0$; (d) $m=1000$, $a=1$, $n=4$.}
\end{figure}

\begin{figure}[htbp]\suppressfloats
\includegraphics[width=15cm]{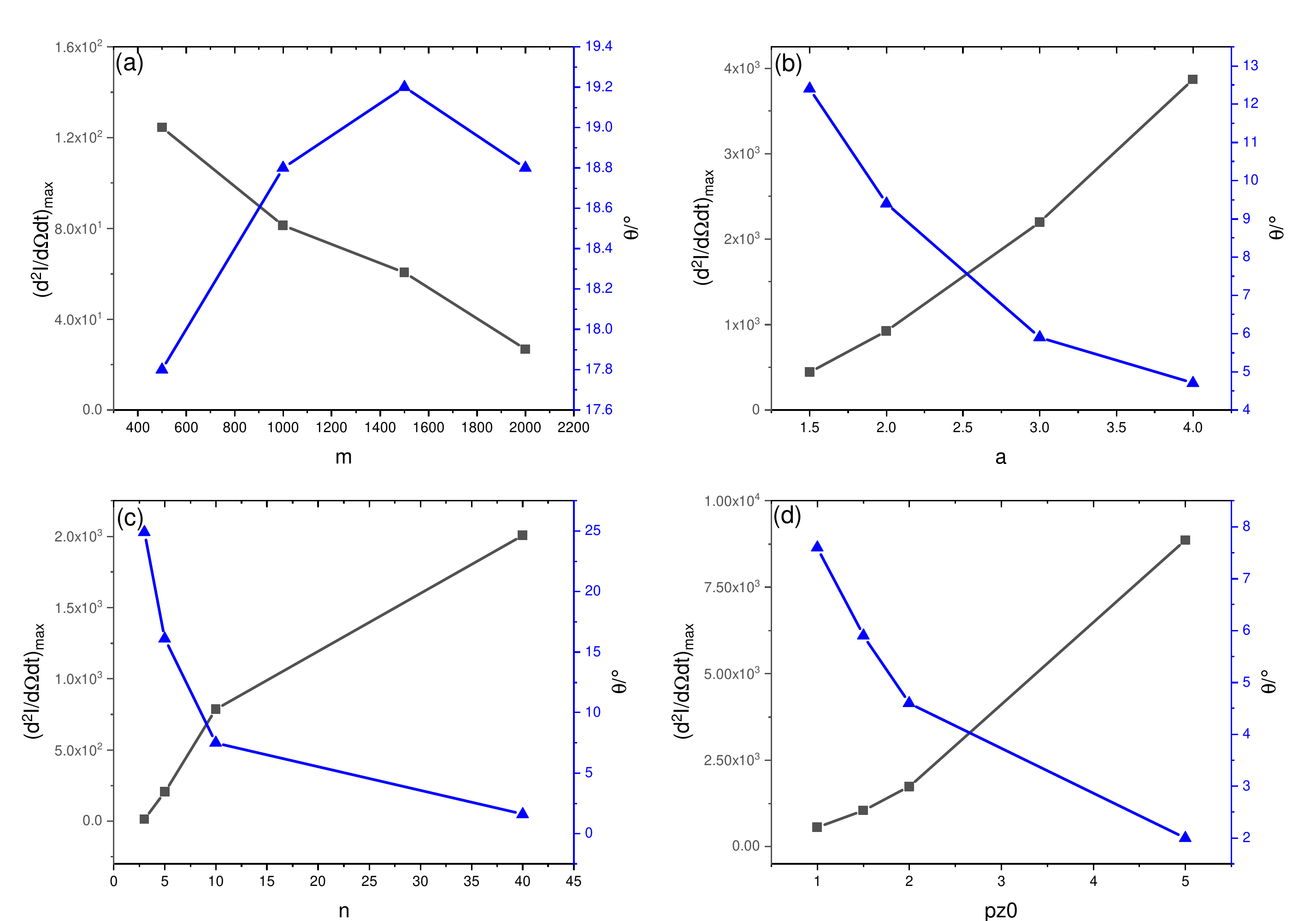}
\caption{\label{Fig12}(colour online) The maximum radiation power and corresponding $\theta$ ($\varphi=0$) change with a set of different $m$, $a$, $n$ and $p_{z0}$. (a) $a=1$, $n=4$, $p_{z0}=0$; (b) $m=1000$, $n=4$, $p_{z0}=0$; (c) $m=1000$, $a=1$, $p_{z0}=0$; (d) $m=1000$, $a=1$, $n=4$.}
\end{figure}

\subsection{Angular distributions with respect to the polar angle $\theta$ when $\varphi=0$}

Furthermore the angular distributions of NTS in the combining field with respect to the polar angle $\theta$ are investigated. And the results for different orders of harmonic radiation $m$, resonance parameter $n$, laser intensity $a$ and initial axial momentum $p_{z0}$ are shown in Fig.\ref{Fig11} and Fig.\ref{Fig12}.
Fig.\ref{Fig11} shows the angular distributions, Fig.\ref{Fig12} shows the change of the maximum radiation power and corresponding $\theta$ with different parameters.
From Fig.\ref{Fig12}(a), we can see that the maximum radiation power decreases when $m$ increases from 500 to 2000, the corresponding $\theta$ varies between $17.8^{\circ}$ and $19.2^{\circ}$. Meanwhile, the $\theta$ corresponding to high radiated power is small which means that the spatial distribution of the radiation approaches the propagation direction of laser. In addition, the corresponding $\theta$ range is relatively wide and is not significantly decreasing with the increasing of $m$. The influence of $a$ on the angular distribution of radiation power is shown in Fig.\ref{Fig11}(b) and Fig.\ref{Fig12}(b), which exhibits that the maximum radiation power increases and $\theta$ decreases with the increasing of $a$, respectively. The larger the $a$ is, the closer the radiation distribution is to the laser propagation direction and the narrower the corresponding $\theta$ range becomes. Moreover, we will take the $1000^{th}$ harmonic radiation and $a=1$ to study the dependencies on other parameters in the following.

Upon the dependence of angular distribution of the radiation power on $n$ and $p_{z0}$, they are examined and shown in Fig.\ref{Fig12}(c) and 12(d). It is found that the maximum radiation power increases and the $\theta$ decreases with the increase of $n$ or $p_{z0}$. The corresponding $\theta$ of the maximum radiation power gradually closes to the laser propagation direction. In fact we find that the larger the $n$ is, the narrower the corresponding $\theta$ range becomes, see Fig.\ref{Fig11}(c), however, the corresponding $\theta$ range of high radiation power are always narrow, see Fig.\ref{Fig11}(d). Specially, it is worthy to note that the radiation distribution is well collimated along the laser-propagation direction ($+z$) when $p_{z0}=5$.

To sum up, the angular distributions with respect to $\theta$ is mainly distributed in the laser propagation direction. Under the conditions that the RRE can be ignored, the radiation power increases as increasing $n$, $a$, and $p_{z0}$, and it increases most obviously as $a$ increases. In particular, if the electron moves fast enough along the $+z$ axis initially, the radiation can be focused on the forward, which results in the possible that the radiation can be collimated in the forward direction under the chosen appropriate set of parameters.

\subsection{The optimum emission direction}

\begin{figure}[htbp]\suppressfloats
\includegraphics[width=15cm]{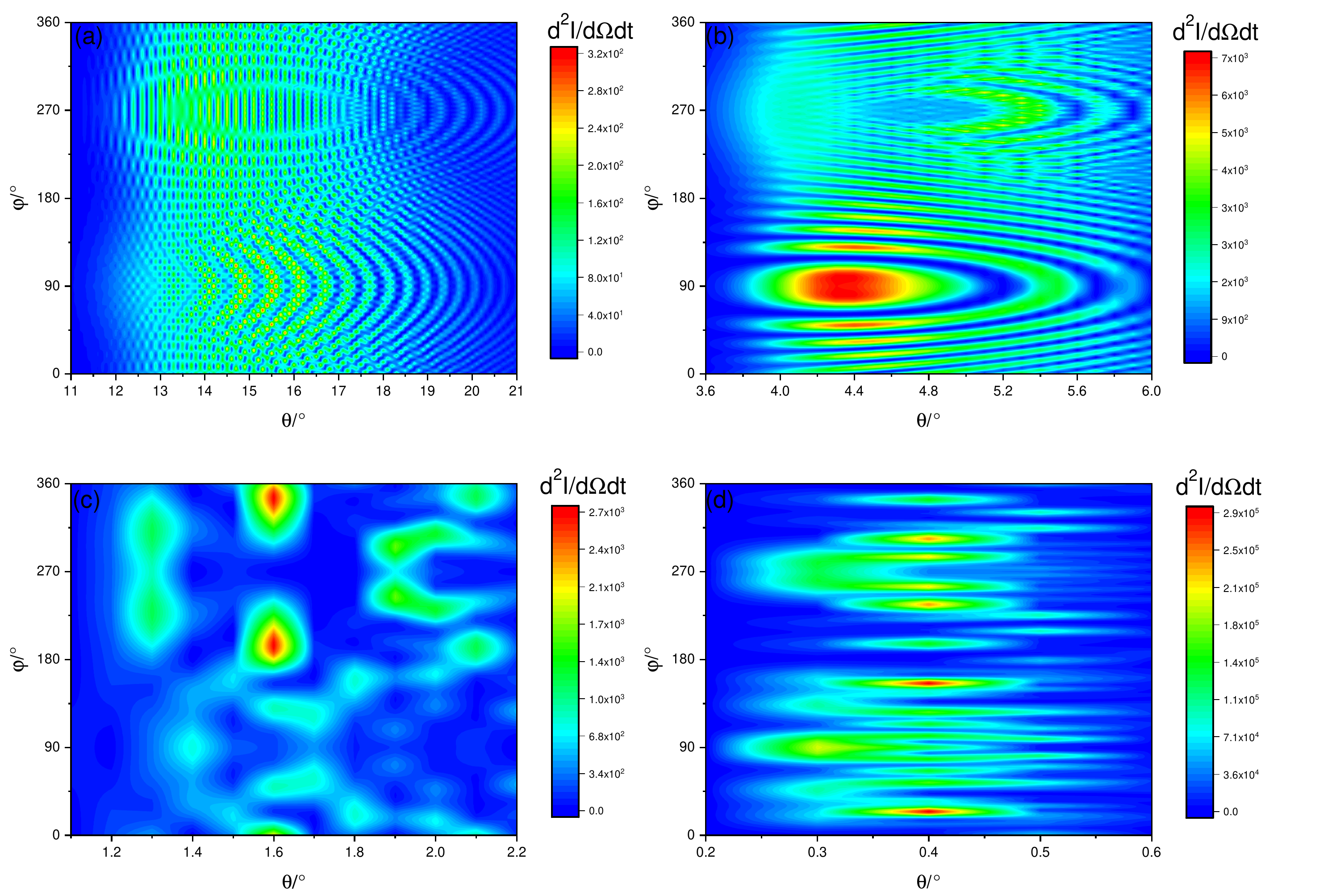}
\caption{\label{Fig13}(colour online) Contour of the spatial distributions of the NTS with respect to the polar angle $\theta$ and the azimuthal angle $\varphi$. (a) $m=1000$, $a=1$, $n=5$, $p_{z0}=0$; (b) $m=1000$, $a=4$, $n=4$, $p_{z0}=0$; (c) $m=1000$, $a=1$, $n=40$, $p_{z0}=0$; (d) $m=1000$, $a=1$, $n=10$, $p_{z0}=10$.}
\end{figure}

From results and discussion mentioned above, we can know that the radiation power with respect to $\theta$ is mainly distributed in the forward direction when $\varphi=0$. The maximum radiation power $(d^{2}I/d\Omega dt)_{max}$, the corresponding $\theta_{max}$ can be obtained. In the full space, the optimum emission direction are investigated, which is expressed by $\theta_{max}$ and $\varphi_{max}$. The involved photon number and radiated power have to change with the emission angles, they are different from the situation of $\varphi=0$.

In Fig.\ref{Fig13}, the contour plotting about the angular resolved pattern and the brightness changing depending on the parameters are shown. For the azimuthal angle $\varphi$, the bright spots of Fig.\ref{Fig13}(c) are concentrated in the range $180^{\circ}$ to $360^{\circ}$ and other figures are concentrated in the range $0^{\circ}$ to $180^{\circ}$. This angular distribution is different from that $xy$ plane which is perpendicular to laser direction. Furthermore, they are all symmetric with respect to $y$ axis ($\varphi=90^{\circ}$ or/and $\varphi=270^{\circ}$) and the results are consistent with previous studies. For $\theta$, they are all in the forward direction. Since the values in the other ranges are small, we focus only on the forward part. It can be see from the pictures, the radiation scattered by electron can be well collimated in the forward ($\theta_{max}\leq5$ degrees) as long as the laser intensity and magnetic field strength are satisfied.

As examples and also the comparable ones to previous study \cite{POP-29-043102}, some results from our research data are listed in Table \ref{tab1} and Table \ref{tab2}. By the way, beside of the case when $\varphi=0$ presented above, here we have studied the full space of $\varphi$, which can provide the more accurate information about the optimal parameters corresponding to the maximum radiation spectra. 

\begin{table}[tbp]
 \setlength\tabcolsep{3pt}
    \caption{\label{tab1}The frequency $\omega$ (normalized by $\omega_{0}$), corresponding radiation energy (normalized by $\textit{e}^{2}\omega_{0}^{2}/4\pi^{2}\textit{c}$ ) and photon number (normalized by $\textit{e}^{2}\omega_{0}/4\pi^{2}\textit{c}\hbar$) per unit solid angle per unit time in the optimum emission direction.}
    \begin{center}
        \begin{tabular*}{15cm}{@{\extracolsep{\fill}}l c c c c c c c c c c}
            \hline\hline
            & $m$ & $a$ & $n$ & $p_{z0}$ & $\theta_{max}(^{\circ})$ & $\varphi(^{\circ})$ & $\omega_{1}$ & $\omega=m\omega_{1}$ & $(d^{2}I/d\Omega dt)_{max}$ & $N_{photon}/d\Omega dt$ \\ \hline
            &$1000$ & $1$ & $5$ & $0$ & $16.1$ & $0$ & $0.1275$ & $1.275\times10^{2}$ & $2.0512\times10^{2}$  & $1.6089\times10^{0}$  \\ \hline
            &$1000$ & $4$ & $4$ & $0$ & $4.7$ & $0$ & $0.1664$ & $1.664\times10^{2}$ & $3.8667\times10^{3}$  & $2.3244\times10^{1}$  \\ \hline
            &$1000$ & $1$ & $40$ & $0$ & $1.6$ & $0$ & $0.0185$ & $1.850\times10^{1}$ & $2.0078\times10^{3}$  & $1.0861\times10^{2}$ \\ \hline
            &$1000$ & $1$ & $10$ & $10$ & $0.5$ & $0$ & $0.0531$ & $5.312\times10^{1}$ & $5.3319\times10^{4}$  & $1.0038\times10^{3}$
            \\ \hline \hline
           & $m$ & $a$ & $n$ & $p_{z0}$ & $\theta_{max}(^{\circ})$ & $\varphi_{max}(^{\circ})$ & $\omega_{1}$ & $\omega=m\omega_{1}$ & $(d^{2}I/d\Omega dt)_{max}$ & $N_{photon}/d\Omega dt$ \\ \hline
            &$1000$ & $1$ & $5$ & $0$ & $14.9$ & $90$ & $0.1344$ & $1.344\times10^{2}$ & $3.2087\times10^{2}$  & $2.3874\times10^{0}$  \\ \hline
            &$1000$ & $4$ & $4$ & $0$ & $4.3$ & $82$ & $0.1759$ & $1.759\times10^{2}$ & $6.9181\times10^{3}$  & $3.9321\times10^{1}$  \\ \hline
            &$1000$ & $1$ & $40$ & $0$ & $1.6$ & $195$ & $0.0185$ & $1.850\times10^{1}$ & $2.7321\times10^{3}$  & $1.4768\times10^{2}$ \\ \hline
            &$1000$ & $1$ & $10$ & $10$ & $0.4$ & $24$ & $0.0639$ & $6.390\times10^{1}$ & $2.8425\times10^{5}$  & $4.4483\times10^{3}$
            \\ \hline \hline
        \end{tabular*}
    \end{center}
\end{table}

From the Table \ref{tab1}, first we can find that $(d^{2}I/d\Omega dt)_{max}$ and $N_{photon}/d\Omega dt$ have increased a few times in case of both optimal $\theta$ and $\varphi$, see the upper part of the table, by comparing with that in case of only optimal $\theta$ when $\varphi=0$, see the lower part of table, however, they have not much differences in the orders of magnitude. It indicates that the optimum emission direction can have a little higher radiation intensity and photon counts.

Concretely, we can clearly see that the $\varphi_{max}$ corresponding to the maximum radiation power is not zero, the corresponding $\theta_{max}$ is also different from that when $\varphi=0$. When $m=1000$, $a=1$, $n=5$, $p_{z0}=0$, the radiation frequency $\omega/\omega_{0}=1.344\times10^{2}$, the maximum radiation power $(d^{2}I/d\Omega dt)_{max}=3.2087\times10^{2}$ ( normalized by $\textit{e}^{2}\omega_{0}^{2}/4\pi^{2}\textit{c}$), photon number $N_{photon}/d\Omega dt=2.3874\times10^{0}$ (normalized by $\textit{e}^{2}\omega_{0}^{2}/4\pi^{2}\textit{c}\hbar\omega_{0}=\textit{e}^{2}\omega_{0}/4\pi^{2}\textit{c}\hbar\approx4\times10^{11}$) , $\theta_{max}=14.9^{\circ}$ and $\varphi_{max}=90^{\circ}$. They are all different from the situation of $\varphi=0$. Therefore, $10^{12}$ photons can be emitted per unit solid angle per second in this situation.
Moreover, the data in Table \ref{tab1} indicate that $10^{12}-10^{15}$ photons can be emitted by the electron per second per unit solid angle.
Therefore, the brightness of the beam at the harmonics we study here maybe has a potential application since the strong radiation of the NTS can reach the frequency range of soft x-ray with photon energy about hundreds of $\mathrm{eV}$, for example, one can get $\omega=1.344\times10^{2}$ when $m=1000$, $a=1$, $n=5$, $p_{z0}=0$ and $\omega=1.759\times10^{2}$ when $m=1000$, $a=4$, $n=4$, $p_{z0}=0$.

For elliptically polarized laser field, $\alpha=0.5$, we have $a=\sqrt{I_{0}[\mathrm{W/cm^{2}}]\lambda{\mathrm{[\mu m]}}/(1.38\times1.25)}$. So, we can estimate the laser intensity as $I_{0}=[{(1.38\times1.25)a^{2}}/{\lambda[{\mathrm{\mu m}}]^{2}}]\times10^{18}\mathrm{W/cm^{2}}\approx1.7\times10^{18}\mathrm{W/cm^{2}}$ when $a=1, \lambda=1\mathrm{\mu m}$. Thus, it is worth noting that the RRE can be ignored when the laser field is weak. Otherwise, the radiation of the electron is very complex and the RRE should be taken into account. Based on the approximation of the external magnetic field $B_{0}$ in Sec II, $B_{0}\approx(1+1/10)\times100\mathrm{MG}=110\mathrm{MG}$ when $n=10,p_{z0}=0$. However, the external magnetic field can be reduced to $B_{0}\approx[(1+1/10)/2\times20]\times100\mathrm{MG}=5.5\mathrm{MG}$ if the the initial axial momentum $p_{z0}=10$ when the resonance parameter keeps still as $n=10$. It indicates that by increasing the initial axial momentum $p_{z0}$ of the electron, the intensity of the magnetic field can be reduced by nearly 20 times, which can be achieved in the experiments now.

\begin{figure}[htbp]\suppressfloats
\includegraphics[width=15cm]{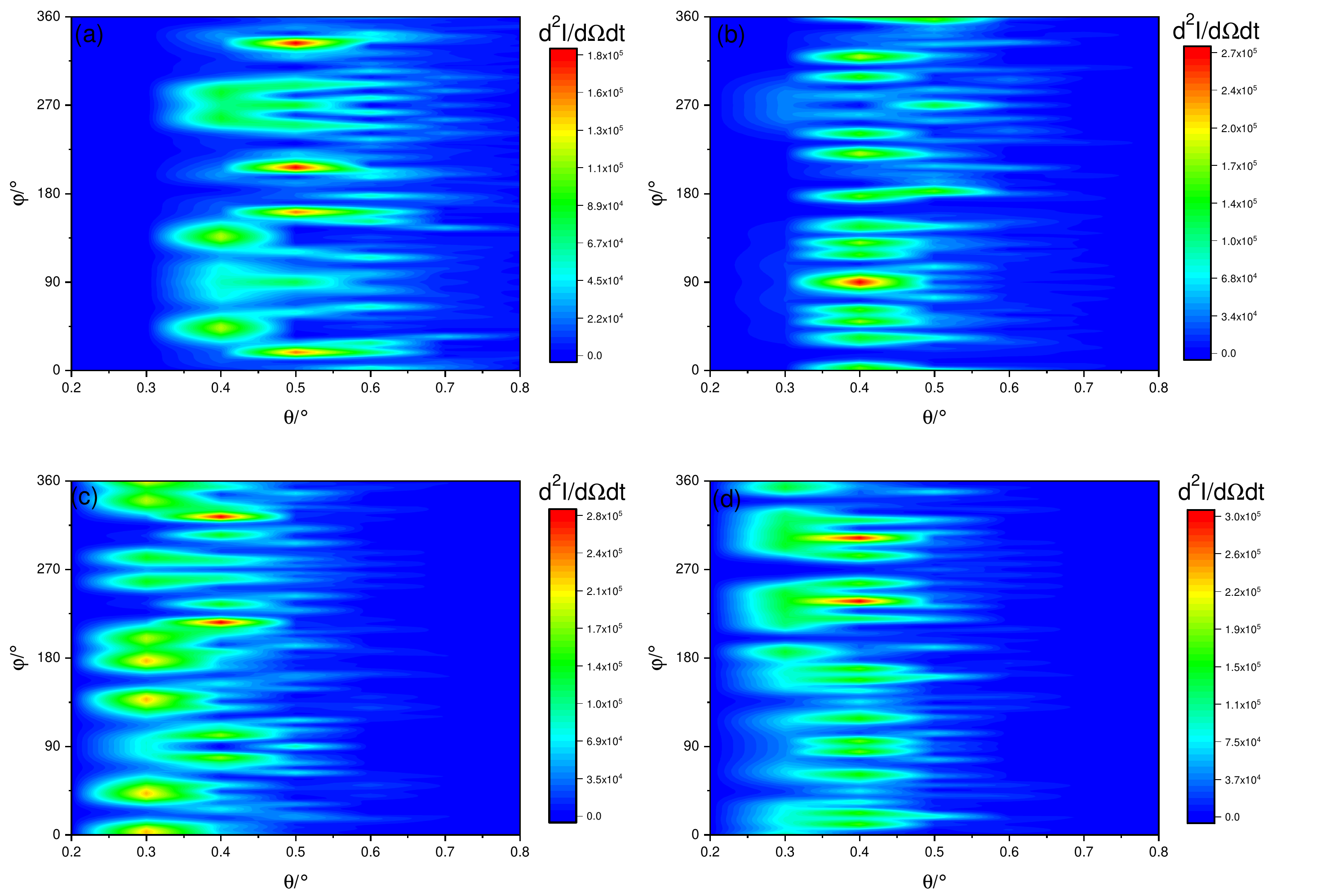}
\caption{\label{Fig14}(colour online) Full spatial distribution of radiated power with different ellipticity $\alpha=0,0.25,0.75,1$. The other parameters are consistent with Fig.\ref{Fig13}(d), $m=1000$, $I_{0}=1.54\times10^{18}\mathrm{W/cm^{2}}$, $n=10$, $p_{z0}=10$.}
\end{figure}

As can be seen from Fig.\ref{Fig14}, the full spatial distribution of the radiated power is symmetric about the $y$ axis and mainly distributed in the forward direction. The position of bright spots change with the increase of $\alpha$ when the other parameters are fixed. The bright spots of $(d^{2}I/d\Omega dt)_{max}$ are concentrated in the range $180^{\circ}$ to $360^{\circ}$ when $\alpha=0.00, 0.75, 1.00$, and others are concentrated in the range $0^{\circ}$ to $180^{\circ}$ when $\alpha=0.25, 0.50$. Moreover, the values of radiated power vary little in orders of magnitude.

\begin{table}[tbp]
 \setlength\tabcolsep{3pt}
    \caption{\label{tab2}The comparison of spatial distribution angle, fundamental frequency $\omega_{1}$ (normalized by $\omega_{0}$), photon number (normalized by $\textit{e}^{2}\omega_{0}/4\pi^{2}\textit{c}\hbar$) per unit solid angle per unit time and photons brightness $B$ corresponding to the maximum radiated power at different ellipticity $\alpha$. The other parameters are $m=1000$, $I_{0}=1.54\times10^{18}\mathrm{W/cm^{2}}$, $n=10$, $p_{z0}=10$.}
    \begin{center}
        \begin{tabular*}{16cm}{@{\extracolsep{\fill}}l c c c c c c c c c}
            \hline\hline
            & $\alpha$ & $\theta_{max}(^{\circ})$ & $\varphi_{1}(^{\circ})$ & $\varphi_{2}(^{\circ})$ & $\omega_{1}$ & $\omega=m\omega_{1}$ & $(d^{2}I/d\Omega dt)_{max}$ & $N_{photon}/d\Omega dt$  & $B$ \\ \hline
            & $0.00$ & $0.5$ & $207$ & $333$ & $0.0662$ & $6.620\times10^{1}$ & $1.7867\times10^{5}$  & $2.6989\times10^{3}$ & $5.63\times 10^{16}$\\ \hline
            & $0.25$ & $0.4$ & $90$ & $90$ & $0.0681$ & $6.810\times10^{1}$ &
            $2.7001\times10^{5}$  & $3.9649\times10^{3}$ & $8.05\times 10^{16}$ \\ \hline
            & $0.50$ & $0.4$ & $24$ & $156$ & $0.0639$ & $6.390\times10^{1}$ & $2.8425\times10^{5}$  & $4.4483\times10^{3}$ & $9.62 \times 10^{16}$\\ \hline
            & $0.75$ & $0.4$ & $216$ & $324$ & $0.0622$ & $6.220\times10^{1}$ & $2.7609\times10^{5}$  & $4.4387\times10^{3}$ & $9.86\times 10^{16}$\\ \hline
            & $1.00$ & $0.4$ & $238$ & $302$ & $0.0618$ & $6.180\times10^{1}$ & $2.9856\times10^{5}$  & $4.8311\times10^{3}$ & $1.08\times 10^{17}$
            \\ \hline \hline
        \end{tabular*}
    \end{center}
\end{table}

The photons brightness of the optimum emission direction of the nonlinear Thomson scattering in combining field with a general elliptical polarization can be estimated as follows. It can be defined as the phase space density of the photon flux, which is given by (refer to Ref.\cite{PRE-48-3003} and Ref.\cite{POP-29-043102})
\begin{equation}
\label{eq21}B\left[\text { photons } / ~\mathrm{s}  ~\mathrm{ mm }^{2} ~\mathrm{mrad}^{2}\right] \simeq 1.40 \times 10^{9} N_{e} \frac{ c[\mathrm{\mu m/s}]}{2.998\times10^{14}} \frac{\lambda_{0}[\mathrm{\mu m}]}{r_{0}^{4}[\mathrm{\mu m}]}\left(\frac{\Delta \omega}{\omega}\right) \frac{\mathrm{d}^{2} I}{\omega^{2} d\Omega dt},
\end{equation}
where $N_{e}=n_{e} \sigma_{0} L_{p}$ is the electron number for the plasma interacting with the laser pulse, $n_{e}$ is the electron density, the bandwidth $(\mathrm{BW})$ $\Delta \omega / \omega \simeq 0.1 \%$, $\sigma_{0}=\pi r_{0}^{2} / 2$ is the laser cross section and $L_{p}=2 \pi r_{0}^{2} / \lambda_{0}$ is the laser-electron interaction distance, $r_{0}$ is the laser spot size.

Their spatial distribution angle, the radiation frequency $\omega$, photon number $N_{photon}/d\Omega dt$ and photons brightness $B$ corresponding to the maximum radiated power $(d^{2}I/d\Omega dt)_{max}$ of polarized laser with different $\alpha$ under the same laser intensity are compared in Table \ref{tab2}. Note that beside the parameters mentioned in Table \ref{tab2}, the dilute plasma density $n_{e}=10^{20} \mathrm{~cm}^{-3}$ is adopted for getting the data.

We can see that $\theta=0.4$ when $\alpha=0.25,0.50,0.75,1.00$, and the corresponding $\omega_{1}$ decrease with the increase of $\alpha$ when $\theta$ are fixed. Now let us look at the photon brightness of the high-order harmonics $m=1000$ in the optimum emission direction from Table \ref{tab2}. By considering the bandwidth, they are about $5.63 \times 10^{19}$, $8.05 \times 10^{19}$, $9.62 \times 10^{19}$, $9.86 \times 10^{19}$ and $1.08 \times 10^{20}$ (in unit of $\text{ photons } / ~\mathrm{s}  ~\mathrm{ mm }^{2} ~\mathrm{mrad}^{2}~0.1 \%\mathrm{BW}$) from the top to bottom, respectively. As a comparison with that of \cite{POP-29-043102}, therefore, it is found that in most cases our results about the emission photon frequency, their counts and brightness are better, while an unavoidable cost is to need an external magnetic field in our study beside of almost the same laser field.

On the other hand, under the same laser intensity in current study, the radiated photons brightness of the circularly polarized laser is the best, and that of the elliptically polarized one is better than the linear one. In addition, the maximum radiated power $(d^{2}I/d\Omega dt)_{max}$ and the photon number per unit solid angle per unit time of circularly polarized laser are better than that of elliptically polarized laser, and elliptically polarized laser is better than linearly one. Under the elliptically polarized laser, the $(d^{2}I/d\Omega dt)_{max}=2.8425\times10^{5}$ and $N_{photon}/d\Omega dt= 4.4483\times10^{3}$ when $\alpha=0.5$ are the highest. So, it is of great interest that we study the angular distributions of nonlinear Thomson scattering in combining field with a general elliptically polarized laser and a background magnetic field.

\section{conclusion and discussion}

In summary, the angular distributions of NTS are studied in detail by an electron moving in combining elliptically polarized laser and magnetic field.

Firstly, the dependence of nonlinear Thomson scattering on ellipticity $\alpha$ are investigated, such as the electrons trajectories, the fundamental frequency, the maximum radiation power in spatial distribution and the corresponding spatial angle. We find that the electron experiences a helically periodic motion in the combining field. As $\alpha$ increases, the electron moves like drawing circles around and the shapes of the trajectories remain unchanged for different $\alpha$. But it can be seen that the electron trajectory drift displacement and maximum oscillation amplitude have increased. The fundamental frequency $\omega_{1}$ decreases with the increase of $\alpha$, polar angle $\theta$, resonance parameter $n$, laser intensity $I_{0}$ or $a$ and initial axial momentum $p_{z0}$. Moreover, $\omega_{1}$ will increase monotonically with the increase of initial phase $\eta_{in}$ in the range $0$-$\pi/2$ and decrease in the range $\pi/2$-$\pi$, it has a maximum value when $\eta_{in}=\pi/2$. So, $\omega_{1}$ changes periodically with the change of $\eta_{in}$ and the period is $\pi$.

For the maximum radiation power in spatial distribution and corresponding spatial angle, we find that they have nonlinear characteristics with the change of $\alpha$, and our main concern is the overall trend. In $xy$ plane when $\theta=\pi/2$, the maximum radiation power decreases with the increase of $\alpha$, $n$, $I_{0}$ and $p_{z0}$, and its value decreases at higher order harmonics. In most cases, the corresponding azimuthal angle approaches $\varphi=270^{\circ}$ as $\alpha$ increases. In $zx$ plane when $\varphi=0$, the maximum radiation power increases with the increase of $\alpha$, $n$, $I_{0}$ and $p_{z0}$, and its value also decreases at higher order harmonics. However, we also find that the elliptically polarized laser is better than the circularly polarized laser for some parameter sets. It means that elliptically polarized laser is the best in some cases and $\theta$ decreases with the increase of $\alpha$, $n$, $I_{0}$ and $p_{z0}$.

Secondly, the angular distribution with respect to $\varphi$ when $\theta=90^{\circ}$ presents interesting shape which is strictly symmetric. They are symmetric about the $y$ axis whatever $m$, $n$, $a$ and $p_{z0}$ are. And the radiation power is stronger when $\varphi$ is in the range of $180^{\circ}\sim360^{\circ}$. Furthermore, the radiated power per solid angle does not decrease monotonously with the increase of $n$ or $a$, and the maximum radiation power can be obtained by choosing appropriate parameters. However, it decreases appreciably and gradually uniformly distributed in the $xy$ plane as the initial momentum increases when $\theta=\pi/2$. Nextly, the angular distribution of radiation with respect to $\theta$ when $\varphi=0^{\circ}$ is well collimated in the forward direction. The larger the $n$, $a$ and $p_{z0}$ are, the closer the radiation distributes to the laser-propagation direction and the larger the maximum radiation power will be.

Accordingly, it is possible that the radiation can be concentrated in the forward direction if an appropriate set of parameters is chosen. We find that the angular distribution with respect to $\varphi$ when  $\theta=90^{\circ}$ and the angular distribution with respect to the $\theta$ when $\varphi=0^{\circ}$ is different. Finally, the best angle of emission $\theta_{max}$ and $\varphi_{max}$ can be accurately figured out and the high frequency part of the radiation can reach the range of XUV and x-ray. In the best emission angles, the number of photons radiated by electron is approaching $10^{12}-10^{15}$ and its photons brightness can reach $10^{20}$$(\text { photons } / ~\mathrm{s}  ~\mathrm{ mm }^{2} ~\mathrm{mrad}^{2}~0.1 \%\mathrm{BW})$. Photons of different frequencies can be radiated along different directions by the electron via NTS. Under the same laser intensity, the radiated photons brightness of the circularly polarized laser is the best, and that of the elliptically polarized one is better than the linear one. With the elliptically polarized laser, the $(d^{2}I/d\Omega dt)_{max}=2.8425\times10^{5}$ and $N_{photon}/d\Omega dt= 4.4483\times10^{3}$ when $\alpha=0.5$ are the highest. It is of great interest that we study the angular distributions of nonlinear Thomson scattering in combining field with a general elliptically polarized laser and a background magnetic field.

While the fixed ellipticity is used for the second-stage numerical study, we believe that the similar treatment can be extended to case of an arbitrary elliptically polarized laser field since the analytical expressions of momenta and displacement of electrons in the combining field are got in the first-stage theoretical description. Meanwhile, the conclusions about azimuthal angle symmetry and polar angle dependence of radiation spectrum which are mentioned above in this paper are general.

Certainly the dependence of the studied problem on the ellipticity is worthy to study for the sake of completeness, which beyond the scope of this paper. Importantly, for practical applications, we think it is particularly important to choose the parameter to produce the photons with specified frequencies that we need. It can be used as a reference to help the experiment researcher to obtain high quality high intensity radiation at the optimal spatial distribution.

\begin{acknowledgments}
This work was supported by the National Natural Science Foundation of China (NSFC) under Grant No.11875007 and No.11935008. The computation was carried out at the High Performance Scientific Computing Center (HSCC) of the Beijing Normal University.
\end{acknowledgments}

\end{document}